\documentclass[aps,pra,superscriptaddress,footinbib,twocolumn]{revtex4-2}

\usepackage{dcolumn}
\usepackage{bm}
\usepackage{float}
\usepackage[T1]{fontenc}
\usepackage{color}
\usepackage{graphicx}
\usepackage[american]{babel}
\usepackage{ulem}

\usepackage{amsmath,amsthm,amsfonts,amssymb,amscd}
\usepackage{indentfirst}
\usepackage{float}
\usepackage[dvipsnames]{xcolor}
\usepackage[colorlinks=true,citecolor=ForestGreen,linkcolor=Mulberry,urlcolor=BlueViolet,hyperindex]{hyperref}
\usepackage{mathtools}
\usepackage{bbold}
\usepackage{units}
\usepackage{soul}
\usepackage{ulem}
\usepackage{xcolor}
\usepackage{caption}
\usepackage{subcaption}
\usepackage{ragged2e}
\DeclareCaptionFormat{justified}{\justifying#1#2#3\par}
\newcommand{\ket}[1]{\vert#1\rangle}

\newcommand{\vcq}{University of Vienna, Faculty of Physics, Vienna Center for Quantum Science and Technology (VCQ) and Research platform TURIS, Boltzmanngasse 5, 1090 Vienna, Austria}
\newcommand{\cdl}{Christian Doppler Laboratory for Photonic Quantum Computer, Faculty of Physics,  University of Vienna, 1090 Vienna, Austria}
\newcommand{\iqoqi}{Institute for Quantum Optics and Quantum Information (IQOQI) Vienna, Austrian Academy of Sciences, Boltzmanngasse 3, Vienna, Austria}

\begin{document}

\title{
Experimental Quantum State Certification by Actively Sampling Photonic Entangled States
}

\author{Michael Antesberger}
\thanks{These two authors contributed equally to this work.}
\affiliation{\vcq}

\author{Mariana M. E. Schmid}
\thanks{These two authors contributed equally to this work.}
\affiliation{\vcq}

\author{Huan Cao}
\email{huan.cao@univie.ac.at}
\affiliation{\vcq}
\affiliation{\cdl}

\author{Borivoje Daki\'{c}}
\affiliation{\vcq}

\author{Lee A. Rozema}
\affiliation{\vcq}

\author{Philip Walther}
\email{philip.walther@univie.ac.at}
\affiliation{\vcq}
\affiliation{\cdl}
\affiliation{\iqoqi}


\begin{abstract}
Entangled quantum states are essential ingredients for many quantum technologies, but they must be validated before they are used. 
As a full characterization is prohibitively resource-intensive, recent work has focused on developing methods to efficiently extract a few parameters of interest, in a so-called verification framework.
Most existing approaches are based on preparing an ensemble of nominally identical and independent (IID) quantum states, and then measuring each copy of the ensemble.
However, this leaves no states left for the intended quantum tasks and the IID assumptions do not always hold experimentally.
To overcome these challenges, we experimentally implement quantum state certification (QSC) proposed by Gocanin \textit{et al.}, which measures only a subset of the ensemble, certifying the fidelity multiple copies of the remaining states.
We use active optical switches to randomly sample from sources of two-photon Bell states and three-photon GHZ states, reporting statistically-sound fidelities in real time without destroying the entire ensemble.
Additionally, our QSC protocol removes the assumption that the states are identically distributed (but still assumes independent copies), can achieve close $N^{-1}$ scaling, in the number of states measured $N$, and can be implemented in a device-independent manner. 
Altogether, these benefits make our QSC protocol suitable for benchmarking large-scale quantum computing devices and deployed quantum communication setups relying on entanglement in both standard and  adversarial situations.
\end{abstract}

\maketitle


Quantum technology uses entangled states as resources to implement tasks with an efficiency or security that cannot be accomplished with only classical resources \cite{QKD_review_Gisin,Deutsch_1998}.
However, before using an entangled state for a given task the experimentally produced state must be verified. Traditionally, this is done by preparing an ensemble of $N$ identical and independently distributed (IID) states and measuring each state. It is widely appreciated that learning the complete quantum state---i.e. performing quantum state tomography~\cite{kwiat_tomography_2001}---is a challenging task, requiring $N$ to increase exponentially with the system size.
This has led to a variety of resource efficient characterization methods, including adaptive state tomography \cite{Mahler2013Adaptive,chapman2016experimental,qi2017adaptive}, compressed sensing \cite{gross2010quantum}, direct fidelity estimation \cite{flammia2011direct}, cross-verification \cite{greganti2021cross,zhu2022cross} and 
quantum state verification (QSV) \cite{hayashi2009group,verification_prx_2018,pallister2018optimal,zhu2019optimal,li2020optimal,hayashi2015verifiable,fujii2017verifiable,hayashi2018self,dimic2018, saggio2019}.
In all of these approaches one measures every copy of the initially prepared ensemble, leaving no states for use in a further experiment.
One must therefore assume that the source operates identically during the characterization and operation phases.
Here, we circumvent this problem by experimentally implementing a \textit{quantum state certification} (QSC) protocol proposed by Gocanin \textit{et al.} \cite{gocanin2022}. 

In QSC only a subset of the ensemble is measured, allowing one to certify some property of the remaining states.
In more detail, we imagine a ``verifier'' who randomly extracts a subset of the initial ensemble of \textit{physical states} and a ``user'' who receives the remaining states.
The verifier performs QSV on their sub-ensemble, allowing her to issue a certificate to the user that his remaining states are close to a promised \textit{target state}.
More precisely, QSC considers a set of $N$ independent (but not necessarily identically distributed) physical states.
The verifier then extracts a random subset of $\mu N$ states, measuring each one.
QSC then answers the question ``\textit{With what confidence can we conclude that the remaining subset of $(1-\mu)N$ physical states have an average fidelity of at least $95 \%$ with the target state?}''

Importantly, QSC is experimentally accessible, meaning that all measurements can be made locally and the post-processing complexity is low.
This is because it is based on QSV which can estimate the fidelity with the optimal $N^{-1}$ scaling \cite{zhang2020,eisert2020quantum}.
While much work on state characterization operates in a trusted, device dependent scenario, there is a growing need to perform device-independent (DI) state characterization with untrusted measurement devices \cite{han2021optimal,zhu2019efficient,zhu2019general}, for applications such as blind quantum computing \cite{hayashi2015verifiable,fujii2017verifiable,barz2012demonstration} and quantum cryptography \cite{zhang2022device,yin2020entanglement}. 
Our work is based on the proposal Gocanin \textit{et al.} \cite{gocanin2022}, which introduces DI-QSV and QSC, allowing us to experimentally realize efficient quantum state certification.
While in our experiment we do not close the loopholes necessary to claim device independence, we provide a detailed discussion of how to extend our experiment to a device-independent setting.


\begin{figure*}[t!]
    \centering
    \includegraphics[width=\linewidth]{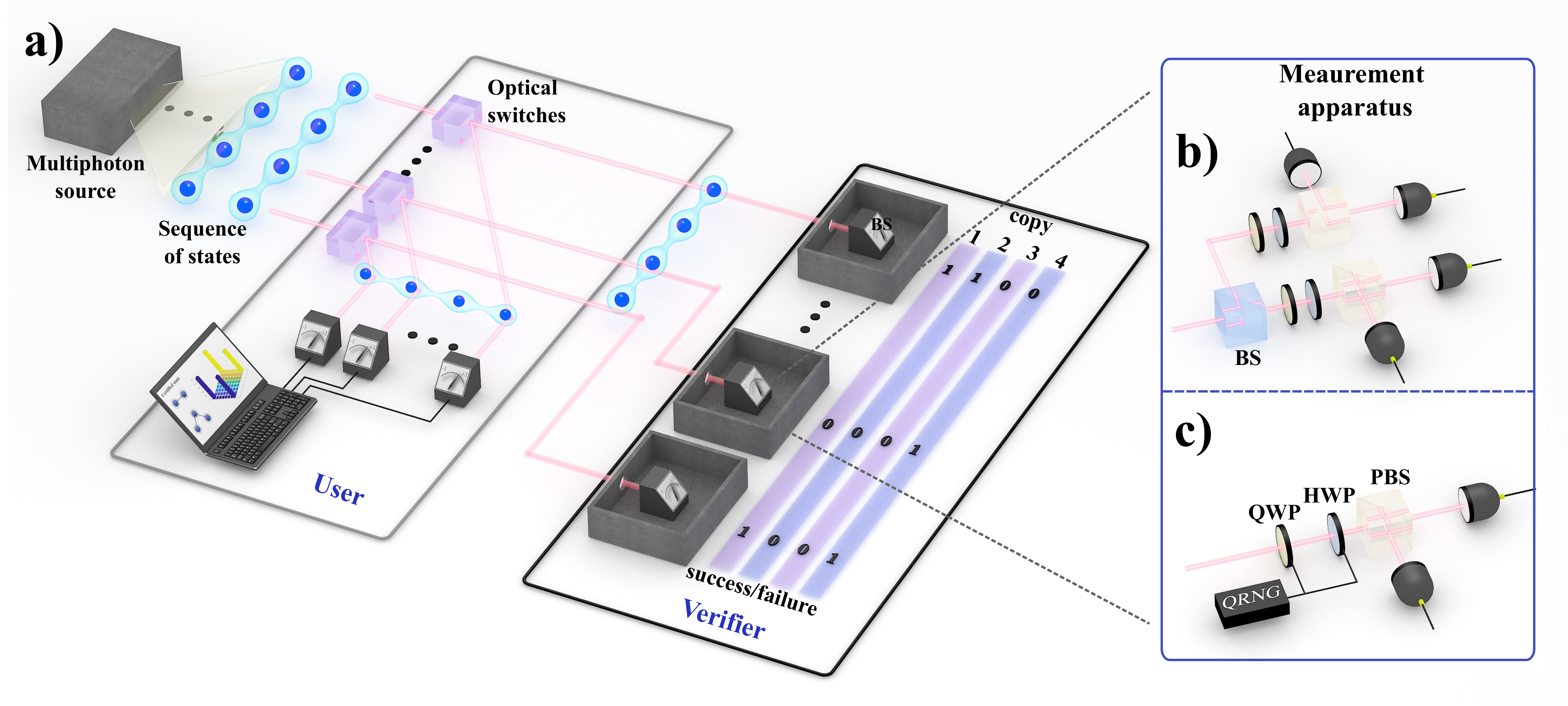}
    \caption{\textbf{Experimental Quantum State Certification.}
    a) Sequences of $M$-photon states are spontaneously produced and sent to $M$ trusted synchronized optical switches.
    The switches periodically alternate between two settings, directing the inputs to the user or verifier.
    We drive the switches faster than the photons are produced. Since the emission time is random, this serves to randomly send each state to the user or verifier.
    Both the user and verifier implement standard polarization measurements on their photons.
    b) Two-photon quantum state verification (QSV) measurement. We randomly select the measurement basis for our two-photon experiment using 50:50 beamsplitters to send each of the $M$ photons to one of two polarization measurements.
    c) Three-photon QSV measurement. For our three-photon experiment, since the three-photon rate is sufficiently low, we randomly select the measurement basis by rotating the waveplates between two settings.  Each subsequent setting is chosen based on the {outcome of} a commercial quantum random number generator (QRNG).
    }
    \label{fig:setup}
\end{figure*}

Photonic Bell states and GHZ states are of extreme importance for a variety of applications.
In particular, two-photon Bell states are an essential primitive for many quantum communication protocols \cite{duan2001long,yin2020entanglement,nadlinger2022experimental}, while three-photon GHZ states and  probabilistic fusion gates can enable the efficient generation of large-scale cluster states  \cite{Gimeno2015threePhoton} for measurement-based quantum computing \cite{walther2005experimental,briegel2009measurement,raussendorf2001one}.
Certifying such resources is essential for these and other applications.
To show the applicability of our QSC protocol, we therefore experimentally implement it for both two-photon Bell states and three-photon GHZ states \cite{pan2012multiphoton,pan2000experimental}.
In our implementation of QSC, some of the produced quantum states are randomly routed to the verifier who performs QSV while the user simultaneously runs his experiment.
The user can either adapt to the current confidence level broadcasted by the verifier or wait until sufficient confidence is achieved.
To realize this deterministically, we use active optical switches \cite{zanin2021fiber}, as outlined in Fig. \ref{fig:setup}.
If the verifier's measurements are successful, the remaining states are certified for the user.
As we use active switches, each individual state is deterministically routed to the user or verifier.
This provides a realistic and practical implementation of QSC, since, from the user's point of view the only effect is a constant reduction in the counting rate, which does not scale with the system size.


\section*{Device Independent Quantum State Certification}
Our implementation of QSC is based on self-testing \cite{mayers2003self}.
While the specifics depend on the target state, the idea of self testing is that certain quantum correlations are almost unique to some target states.
For example, self-testing for a GHZ state works as follows. One attempts to violate a Mermin inequality \cite{mermin1990} with the physical states. 
This can only achieved if the physical states are close to the target state.
Then, if the violation is sufficiently strong, one can place bounds on the average fidelity $\bar{F}$ of the physical states.
We mention that in a DI framework it is not possible to directly assess the fidelity. 
Instead, one uses the extractability $\Xi$, which is equivalent to the fidelity up to local isometries \cite{kaniewski2016analytic,mayers2003self,zhang2018experimentally}. 
{However, for simplicity, in the following, we use fidelity $F$ (or infidelity $\eta=1-F$) in place of the extractability if not otherwise specified.}
Since self-testing is based on estimated probabilities {it usually does not discuss the scaling behaviour for finite $N$.}
\cite{vsupic2020self,mayers2003self}.
To work in this regime, QSV converts self-testing to a non-local game that can only be won by the target state \cite{gocanin2022}. 

To explain QSV, consider a source producing an independent sequence of N states $S=\left\{\sigma_1, \sigma_2, ..., \sigma_N\right\}$, and a target state $\ket{\psi}$ that for now we take to be a three-photon GHZ state.
The Mermin inequality for this target state is \cite{mermin1990}:
\begin{widetext}
\begin{equation}
    \label{eq:mermin_inequality}
    B=\sum\limits_{o_1,o_2,o_3}(-1)^{o_1+o_2+o_3}\left[p(o_1,o_2,o_3|0,0,1)+p(o_1,o_2,o_3|0,1,0)
    +p(o_1,o_2,o_3|1,0,0)- p(o_1,o_2,o_3|1,1,1)\right] \leq 2.
\end{equation}
\end{widetext}
Here the individual terms are probabilities of outcomes to be estimated experimentally.
In more detail, $p(o_1,o_2,o_3|i_1,i_2,i_3)$ are the probabilities to obtain outcomes $o_1$, $o_2$ and $o_3$ when qubits one, two and three are measured with a setting defined by $i_1$, $i_2$ and $i_3$. The outputs $o_i$ are assumed to be binary, taking values zero or one, which leads to the separable state bounds $B\leq 2$.
For an ideal GHZ state, however, the quantum bound is $B_q=4$.
For the Mermin inequality (Eq.\ref{eq:mermin_inequality}), there are four measurement settings: $\{i_1,i_2,i_3\}=\{0,0,1;~0,1,0;~1,0,0;~1,1,1\}$ (defined in the Materials and Methods Eq. \ref{eq:measurements}) each with eight possible outcomes.
For a perfect GHZ state {only a subset of outcomes} will occur with certainty: we group these together, forming the \textit{winning outcomes}.
To make the inequality into a non-local game, one then randomly measures each state of $S$ in one of the four settings, records the number of winning events $n_\mathrm{wins}$, and computes the experimental winning probability as $P_\mathrm{exp}={n_\mathrm{wins}}/{N}$.
The intuition is then that since only states equivalent to the target state can violate the inequality, only they can obtain a $P_{exp}$ close to $1$. 

\begin{figure}[b]
    \centering
    \includegraphics[width=\linewidth, trim=0.2cm 0.cm 1.cm 0.1cm, clip]{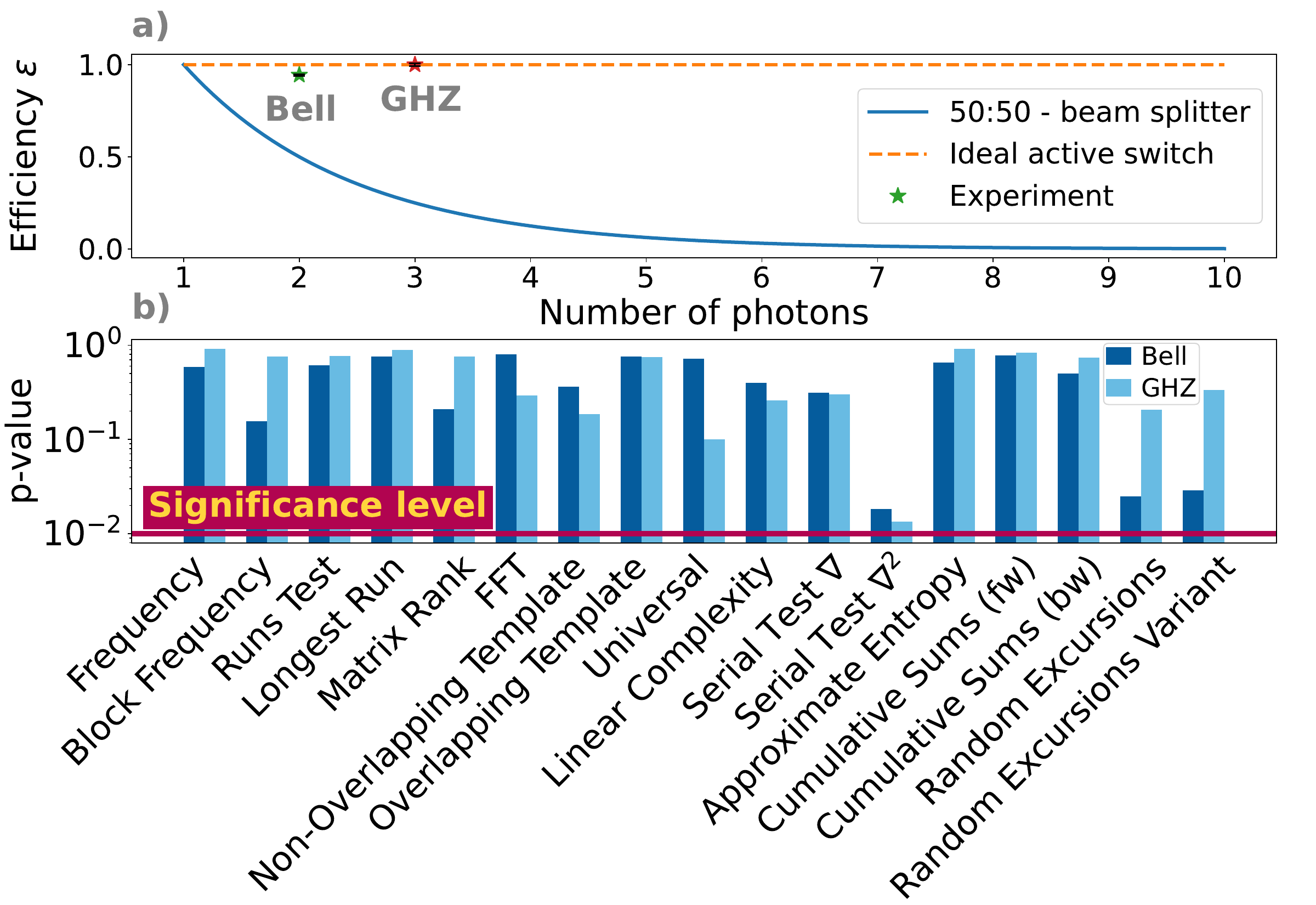}
    \caption{\textbf{Random Sampling Characterization.} (a) Synchronization efficiency. The blue curve plots the maximum synchronization efficiency using probabilistic routing with 50:50 beam splitters versus the number of photons in the state to be certified. The dashed line at $1$ indicates the maximum efficiency attainable with the deterministic switches that we use. Our observed two- and three-photon efficiencies of
    $\epsilon_{2} = 0.9439 \pm 0.0041$, and $\epsilon_{3} =0.9997 \pm 0.0066$, respectively, are plotted as stars. (b) The results of NIST's statistical test suite for random number generators applied to our sampling of the produced state for two-photon Bell states in dark blue, and three-photon GHZ states in light blue. In both cases, the results exceed the significance threshold of $0.01$, indicating that our implemented sampling is consistent with truly random sampling.
    }
    \label{fig::randomness}
\end{figure}

In QSC, the verifier measures $N_\mathrm{ver}=\mu N$ states of the initial $N$ state ensemble, and uses her measurement results to estimate $P_\mathrm{exp}$.
The remaining $(1-\mu)N$ states are then certified.
Previous related work was formulated such that only a single copy was certified: \textit{i.e.} $N-1$ states are measured and only one state is certified {\cite{zhu2019efficient,zhu2019general}}.
Doing so one can completely remove all IID assumptions.
Ref. \cite{gocanin2022}, instead, formulates QSC for any value of $\mu$, allowing multiple states to be certified. This comes at the cost of still requiring the assumption that the subsequently emitted states are independent, but, nevertheless, still removes the identical distribution assumption (see Ref. \cite{gocanin2022} for more details). 
Given that state independence is a very reasonable assumption for parametric photon-pair sources, here we implement multi-copy certification, which may be applicable for a wider class of applications.

To formalize the relation between $P_\mathrm{exp}$ and $\bar{\Xi}$ the average extractability, consider a sequence of $\mu N$ physical states with a reduced average extractability  $\bar{\Xi} \leq 1-\eta$.
In this case, the maximum winning probability is $P_\mathrm{\eta}=P_\mathrm{QM}-c\eta$, where $c$ is a constant depending on the specific Bell inequality \cite{gocanin2022}, and $P_\mathrm{QM}$ is quantum mechanical probability to win the game (which is $1$ for our present example).
If we then observe $P_\mathrm{exp}>P_\mathrm{\eta}$, the physical states likely have an extractablity $\bar{\Xi}>1-\eta$.
However, if $P_\mathrm{exp}$ is close to $P_\mathrm{\eta}$ it is possible that finite statistical fluctuations led to an `accidental' win.
Thus we need to know how likely it is that our observation, $P_\mathrm{exp}>P_\mathrm{\eta}$, could be reproduced by a series of states with a fidelity less than $F=1-\eta$.
This is a classical statistical problem \cite{chernoff1952measure,hoeffding2012collected}.
It can be shown that the probability for a false win after $\mu N$ states are measured is bounded by $\delta \leq e^{-D(P_\mathrm{exp}\parallel P_\mathrm{\eta} ) \mu N}$, where $D(x\parallel y)=x \mathrm{log}(x/y)+(1-x) \mathrm{log}\left[(1-x)/(1-y)\right]$ is the Kullback-Leibler divergence \cite{chernoff1952measure}. 
Based on this, Ref. \cite{gocanin2022} proves that our confidence $C$ that remaining $(1-\mu)N$ states have an  average fidelity is greater than $1-\eta$ is
\begin{equation}
    \label{eq:confidence}
    C\geq 1-\delta = 1-e^{-D(P_\mathrm{exp}\parallel P_\mathrm{\eta} )\mu N},
\end{equation}
which grows to $1$ exponentially fast with $N$.
See the Supplemental Material Sec. G for more details.

\begin{figure*}[t]
    \centering
    \includegraphics[width=\linewidth]
    {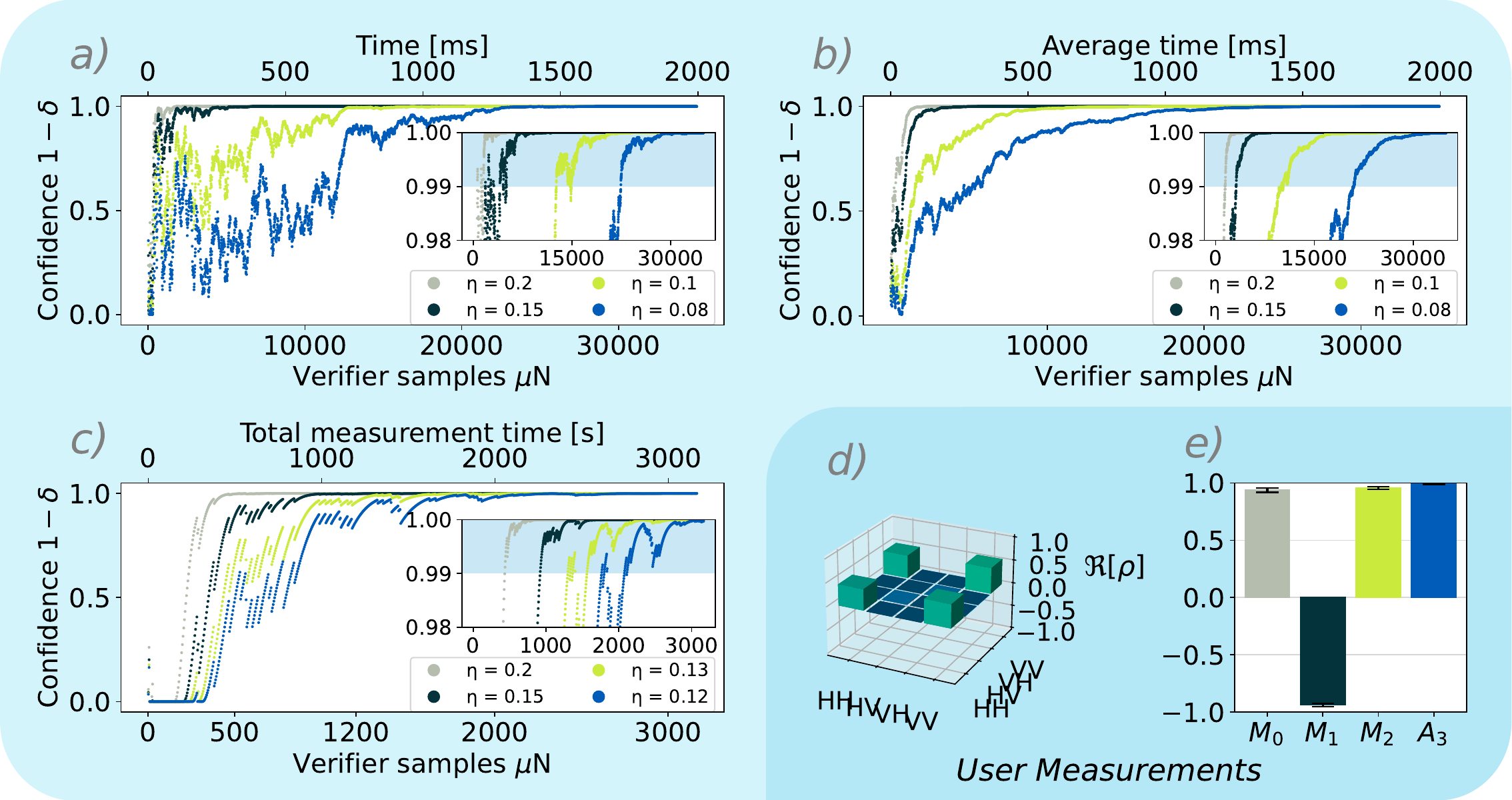}
    \caption{\textbf{Confidence in two- and three-photon quantum state certification.} (a) Confidence versus number of verifier measurements (bottom axis) and measurement time (top axis). The same data set is analyzed for different fidelity $F=1-\eta$ lower bounds. Less stringent bounds (higher values of $\eta$) converge to confidence of $1$ more quickly. The inset shows the high-confidence region, with the shaded area representing a confidence above $0.99$.
    (b) The two-photon measurement presented in (a) is repeated $12$ times and averaged. The data are plotted as in panel (a). 
    (c) Verifier's confidence growth for our three-photon GHZ states, plotted as in panel (a). 
    (d) Real part of the density matrix of our two-photon states measured with standard quantum state tomography, taken by the user concurrently with the verifier's measurements.
    The plotted density matrix has a fidelity of $0.9947 \pm 0.0002$ with the target $\ket{\phi^+}$ Bell state.
    (e) The results of the user's GHZ witness measurement on our three-photon states, yielding a fidelity of $0.9678\pm 0.0055$ with the target GHZ state.
    The labels on the x-axis refer to specific measurement settings, defined in the Supplementary Material.
}
    \label{fig::confidence}
\end{figure*}

The scaling of the confidence that our physical states have an extractability larger than $1-\eta$ depends on the statistical distance between $P_\mathrm{exp}$ and $P_\mathrm{\eta}$ (eq. \ref{eq:confidence}).  
The estimated number of measured states needed to verify this lower bound is 
{$N \geq \frac{\mathrm{ln} \ \delta}{\mathrm{ln}(1-\mu+\mu e^{D(P_\mathrm{exp}\parallel P_\mathrm{\eta})})}$} \cite{gocanin2022}.
Experimentally, instead, we often wish to compute the fidelity from the results from a set of measurement results.
We can also use QSC for this.
It results in two different scaling behaviours related to the specifics of the Bell inequality used to make the non-local game.
In particular, not all states have Bell inequalities that achieve a perfect success probability when translated to a non-local game. 
This happens when quantum mechanics cannot achieve the maximum algebraic bound of the Bell inequality.
For example, for the CHSH inequality the quantum Tirelson bound $B_q=2\sqrt{2}$ is less than the algebraic bound of $4$. 
When this occurs, we also need to consider the ideal winning probability $P_\mathrm{QM}$.
These scenarios arise when some winning outcomes do not occur $100\%$ of time.
We can still define the set of winning outcomes, but now they are given by the most likely outcomes. 
When $p_{QM}<1$, the verifiable infidelity scales as {$\eta \propto O(\frac{\sqrt{\mathrm{ln} \ \delta^{-1}}}{c\mu(1-\mu)\sqrt{N}})$}, i.e. with a sub-optimal $N^{-1/2}$ scaling. This $N^{-1/2}$ scaling behaviour is often called the standard quantum limit. 
However, when $p_{QM}=1$ we can obtain the optimal $N^{-1}$ so-called Heisenberg scaling {$\eta \propto O(\frac{\mathrm{ln} \ \delta^{-1}}{c\mu(1-\mu)N})$}.
{Note that these scaling behaviours are derived in the limit of large $N$, and, as we will see later, in our intermediate regime, they are only approximate.}
Nevertheless, as we will show experimentally, this means that for the two-photon Bell state, using the CHSH inequality we expect to achieve {approximate} $N^{-1/2}$ scaling, but the 3-photon GHZ state with the Mermin inequality can exceed this scaling.

\section*{Experimental Apparatus for Quantum State Certification}

We will now present our experimental implementation of QSC (as proposed by Gocanin \textit{et al}. in Ref. \cite{gocanin2022}) for bipartite and tripartite states, allowing us to demonstrate both scaling behaviors discussed above.
For the two-photon case, we use a type-0 spontaneous parametric down-conversion (SPDC) source in Sagnac configuration \cite{type0_source_Ursin} to generate the Bell-state $\ket{\Phi^+} = \frac{1}{\sqrt{2}}(\ket{HH}_{1,2} + \ket{VV}_{1,2})$, while for the tripartite case, we employ a three-photon Greenberger-Horn-Zeilinger (GHZ) state $\ket{\Psi_{GHZ}} = \frac{1}{\sqrt{2}}(\ket{HHH}_{1,2,3} + \ket{VVV}_{1,2,3})$ produced by two sandwich-like SPDC sources \cite{zhang2015}.
Our three-photon source is a post-selected source, meaning that in order to ensure that the source has generated the correct entangled state we must post-select on three photons arriving at each measurement station (as well as detecting an additional trigger photon in the source).
With respect to a DI implementation of QSC, this opens up our source to the post-selection loophole \cite{Aerts1999TwoPhoton,Larsson_2014,Huang2022Experimental} which cannot be closed with our current methods.
Doing so would require a source which directly generates three photon entanglement \cite{hamel2014direct} with high enough coupling efficiency or a heralded three-photon source \cite{walther2007heralded, cao2024photonic} with sufficiency high heralding efficiency.
We stress that this post-selection loophole is not related to our use of probabilistic sources, which have been used for other DI protocols \cite{shalm2021device}.

For a full description of the photon sources, see the Supplementary Material, {Sec. A and B}.
In both scenarios, we assume that the subsequent states emitted by the source are independent, but we do not need to assume that they are identically distributed.

\begin{figure}
    \centering
    \includegraphics[width=\linewidth]{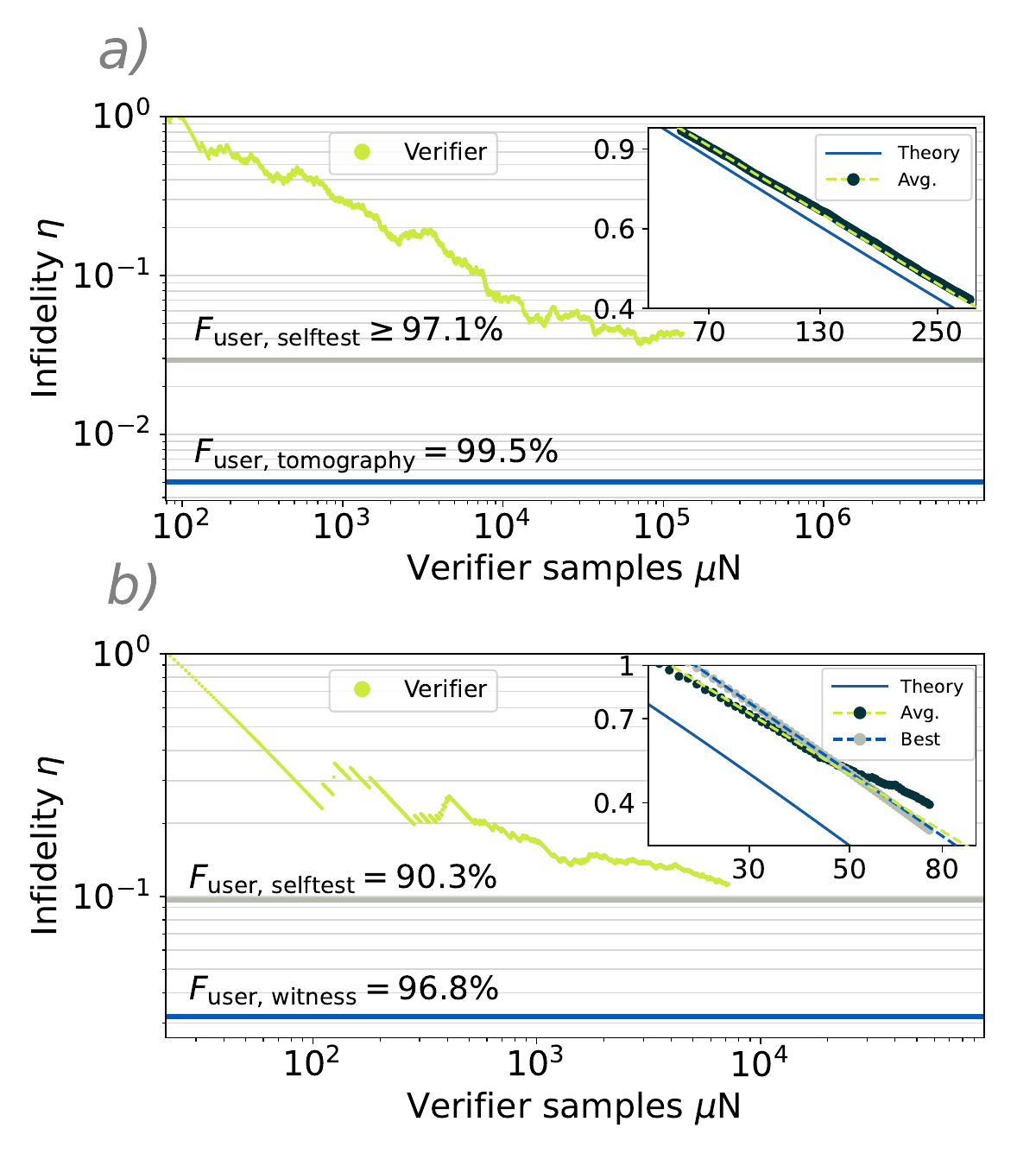}
    \caption{\textbf{Quantum state certification fidelity scaling at $99\%$ confidence.} The estimated infidelities $\eta$ when setting the confidence $C=1-\delta=0.99$ for (a) two-photon Bell states and (b) three-photon GHZ states plotted versus the verifier's measurement number. In both cases, the infidelity $\eta$ certified by the verifier (green points) asymptotically approaches infidelity estimated by the user with self-testing (grey lines).
    The user estimates lower infidelities when using device-dependent methods (blue lines)
    \textbf{Insets:} The insets plot averaged results over smaller sample ranges to estimate the scaling $s$. The dark points are the averaged data, and the dashed green lines are linear fits with slopes of ($s_{\text{Bell, Avg.}} = -0.549$, $s_{\text{GHZ, Avg.}} = -0.801$), respectively. 
    The solid blue line is theory, demonstrating scalings of $s_{\text{Bell, Theory}} = -0.566$, $s_{\text{GHZ and Theory}} = -0.935$.
    In the inset of panel b) we additionally plot the best-case scaling from one data run that achieves scaling of $s_{\text{GHZ, Best}} = -0.907$.
    }
    \label{fig::infidelity}
\end{figure}

To randomly route the multi-photon states between the user and verifier for QSC one could use passive beamsplitters.  While this would work perfectly for a single photon state, it introduces significant loss for $M$-photon states.
For QSC, we need all of the photons to either arrive at the user or verifier; situations where, for example, one photon arrives at the user and another at the verifier count as loss events.
Since with 50:50 beamsplitters each photon is independently routed, all $M$ photons will arrive at either the user or verifier with probability $\epsilon=\left(\frac{1}{2}\right)^M$.
In other words, the photons are not synchronously routed between the user and verifier.
This is the passive synchronization efficiency plotted as the blue line in Fig. \ref{fig::randomness}a).
In a fully DI implementation of QSC, the detection efficiency must remain sufficiently high. This synchronization loss serves to decrease the detection efficiency, making DI-QSC impossible with passive beamsplitters.
Even if a DI implementation is not desired, the loss introduced by sampling with passive elements tends to $0$ exponentially fast, making active elements essential for larger entangled states.

To circumvent this loss, we use $M$ synchronized optical switches (OS), see Fig. \ref{fig:setup}.
Due to the different photon production rates, we use different technologies for our two- and three-photon experiments, {explained in the Materials and Methods Sec. A}.
In any case, in both experiments all $M$ OSs switch between the user and verifier. 
{Both switches are driven with a fixed $50\%$ duty cycle, to achieve $\mu\approx 0.5$, but this can be tuned to set other sampling rates.}
We show below how this can still be used to randomly sample from the states.
To characterize the QSC efficiency for both experiments and show we exceed the passive synchronization efficiency, we measure the two- and three-rates at the user's (verifier's) measurement station $R_\mathrm{use}$ ($R_\mathrm{ver}$), as well as all two- and three-photon events between the two measurement stations $R_\mathrm{cross}$ (nominally, $R_\mathrm{cross}=0$). Then the experimental syncronization efficiency is $\epsilon=(R_\mathrm{use}+R_\mathrm{ver})/(R_\mathrm{use}+R_\mathrm{ver}+R_\mathrm{cross})$. These are the data points shown in Fig. \ref{fig::randomness}a, confirming we have routed the photons with high efficiency.
See the Supplementary Material Sec. C for more details.
Although the total loss in our experiment is too high to close the detection loophole, our synchronization efficiency is high enough ($0.9439 \pm 0.0041$ and $0.9997 \pm 0.0066$ for our two- and three-photon states, respectively) that is would not significantly affect the detection loophole in an otherwise loophole-free experiment. In the Methods Sec. C, we show that also including the insertion loss of the switches used in our two-photon experiments would decrease the overall too much to remain loophole free, while the switches used in our three photon experiment are already sufficient for a loophole-free experiment.

Our OSs sequentially alternate between the user and verifier, which, on its own, does not implement random sampling.
To that end, we make use of SPDCs inherent randomness, (making a device-dependent assumption).
In particular, we drive our OSs much faster than our sources produce photons.
Thus, when a given multi-photon state is randomly emitted the OSs will randomly be in one configuration or the other.
To confirm this, we call an $M$-photon detection at the user (verifier) a `0' (`1'), and generate a bit-string.
We test these bit strings with NIST's statistical test suite for random number generators \cite{bassham2010}.
As elaborated on in the Supplementary Material Sec. D, we obtain an overall confidence of $0.99$ that our produced bit-strings are random. Fig. \ref{fig::randomness}b shows the results of the individual tests.
Although this analysis indicates that we have achieved random sampling, doing so requires us to make a device-dependent assumption that the randomness originates from the source. For a fully DI implementation the switches should be driven randomly, as we discuss in the {Methods Sec. C}.

Once the photons arrive at the verifier's measurement station, she must randomly choose a measurement setting.
For our two-photon and three-photon states we use the measurement sets $M_\mathrm{CHSH}$ and $M_\mathrm{Mermin}$, respectively, which are defined in Eq. \ref{eq:measurements} in the Materials and Methods section.
In both cases, the measurements are implemented with standard methods (waveplates and polarizing beamsplitters), but because of the different photon production rates we randomly sample from the measurements differently (See Fig. 1 panels b and c).

\section*{Experimental Results}

With our QSC apparatus in place, we drive our OSs with 50\% duty cycle pulses so that the verifier takes approximately half of the photons for her measurements.
{In our two-photon measurements, this results in counting rates of {$R_\mathrm{user} \sim 33$ kHz ($R_\mathrm{ver} \sim25$ kHz)} at the user (verifier), resulting a sampling probability of $\mu_\mathrm{Bell}\approx 0.43$ for our two-photon experiment. For the three-photon experiment we have $R_\mathrm{user} \sim 0.3$ Hz ($R_\mathrm{ver} \sim 0.25$ Hz) at the user (verifier), yielding $\mu_\mathrm{GHZ}\approx 0.45$.}
Note that we have defined $\mu=R_\mathrm{ver}/(R_\mathrm{user}+R_\mathrm{ver})$.
As mentioned above, QSC can be performed for any value of $\mu$. Here we set it close to $0.5$ so that the user and verifier measure approximately the same number of measurement results to ease their comparison.
While the verifier implements QSV, the user operates his measurements independently.
We implement two different characterizations for the user and check their consistency with the verifier.
First, the user implements standard device-dependent characterizations. For our two-photon source, the user performs full quantum state tomography, finding a fidelity of $0.9947 \pm 0.0002$ with the target Bell state (see Fig. \ref{fig::confidence}d).
Given the lower three-photon count rate and the larger number of required measurements, for three-photons the user instead uses a GHZ witness \cite{huang2011experimental,yao2012observation} to estimate a fidelity of $0.9678\pm 0.0055$ (see Fig. \ref{fig::confidence}e).
We also perform standard DI self-testing with the user's measurement device, finding a fidelity of $0.971 \pm 0.005$ and $0.9032\pm 0.0066$, for our two- and three-photon states, respectively.
See the Supplementary Material Sec. F for more details.
The discrepancy between the device-dependent and DI techniques is well-known, arising because it is, in general, easier to place tighter bounds in a device-dependent scenario \cite{bancal2014device}. 

While the user performs his measurements the verifier implements QSV on her subset.
To do so, she simply records the number of winning and losing events and computes the winning probability $P_\mathrm{exp}$ as a function of the number of measurements.
$P_\mathrm{exp}$ for both bipartite and tripartite scenarios is provided in the Supplementary Material Sec. E.
To use $P_\mathrm{exp}$ to certify the physical states the verifier sets a minimum fidelity $1-\eta$, and uses Eq. \ref{eq:confidence} to compute her confidence in that lower bound.
In Fig. \ref{fig::confidence}a), we plot this confidence versus the number of measurements made by the verifier $\mu N$ for a range of infidelities from $\eta=0.08$ to $\eta=0.2$ for the two-photon states.
Fig. \ref{fig::confidence}c) shows the same data for our three-photon states for a 
range of $\eta=0.12$ to $\eta=0.2$.
For each data set, we observe a series of rapid increases and sharp drops.
As $\mu N$ increases the drops become less pronounced and the confidence tends to $1$.
The periods of increasing confidence are caused by successive winning events, while the drops are caused by losses.
{We also stress the measurement times for these data sets.  In the two-photon data, in less than $\approx 1$ second, the verifier reports a fidelity above $0.9$ with a confidence $\gg 0.99$, demonstrating the applicability of our methods.
Given our lower three-photon rates, we achieve similar values in $\approx 30$ minutes for the GHZ states.
Nevertheless, in both scenarios, the user and verifier can operate concurrently, with only a constant decrease in the user's count rate. }

To smooth out the statistical fluctuations in the two-photon data, we perform $12$ rounds of QSV with $35000$ samples and average the results. This results in the curves in Fig. \ref{fig::confidence}b).
We identify the minimal sample number required to reach $99\%$ confidence level. The insets in panels a) and b) present zoomed-in plots of the high-confidence region, and the blue-shaded areas denote the area above $99\%$. 
From the inset in b), we can estimate the number of samples need to certify infidelities below 
{$\eta=\left\{0.2,0.15,0.1,0.08\right\}$}, finding 
{$\left\{1420, 3106, 10019, 20982\right\}$} samples are needed, respectively, for our two-photon Bell states.

The main difference between the two-photon and three-photon data sets is related to the different values of $P_\mathrm{QM}$.
For our two-photon CHSH results, $P_{QM}=\frac{2 + \sqrt{2}}{4}\approx0.85$.
For the Mermin inequality, however,  $P_\mathrm{QM}=1$, leading to fewer failed events and, thus, smoother data in a single experimental run, obviating the need to average to estimate the required number of measurements. 
Overall, for the three-photon experiment, we have a success probability $P_\mathrm{exp}\sim 0.97$, which leads to $\left\{433,919,1562,2111\right\}$ samples to certify infidelities $\eta=\left\{0.2,0.15,0.13,0.12\right\}$ with $99\%$ confidence, respectively.

{As discussed above, different values of $P_{QM}$ lead to different scaling behaviours for our two- and three-photon measurements.} 
To observe this, we again perform QSC, now directly comparing the verifier's claims on the infidelity to that measured by the user.
For the verifier, we then instead fix the confidence level $1-\delta$ to $0.99$, and numerically solve Eq. \ref{eq:confidence} for the infidelity $\eta$.
This then represents the lowest DI-infidelity that our sequence of measurements is consistent with at a $99\%$ confidence level.
Based on this, we plot $\eta$ versus the number of verifier measurements $\mu N$ in Fig. \ref{fig::infidelity}. Therein, panel a) shows the two-photon data, and panel b) the three-photon data.
{The bounds in both plots represent two alternative methods to estimate the fidelity discussed above.} 
{The bounds indicated by the blue lines come from device-dependent measurements (quantum state tomography for two-photon case, and GHZ witness \cite{huang2011experimental,yao2012observation} for our three-photon states).
On the other hand, the bounds in grey come from DI self-testing (Supplemental Material Sec. F). 
All of these bounds are measured by the user.
In both cases, when sufficient data is acquired QSC converges to the DI self-testing bound.
Importantly, these data show that our QSC protocol is consistent: the infidelities the user measures (with DD or DI methods) are all lower than the infidelities reported by the verifier.}

Finally, we analyze the scaling of the infidelity as a function $\mu N$.
To this end, we average several repetitions of the experiments over a fixed sample range. 
The resulting averaged estimated infidelities, are shown in the insets of Fig. \ref{fig::infidelity}.
For the bipartite case, we take 1398 repetitions for $N$ from $1$ to $300$ 
(points in Fig. \ref{fig::infidelity}a inset), performing a linear fit in the log-log scale and obtaining a slope of $-0.549$ (dashed green line in the same inset). This means that the scaling is $\eta\propto N^{-0.549}$.
This already exceeds the predicted asymptotic scaling of $N^{-0.5}$.
While this may seem surprising, both the Heisenberg limit scaling and the standard quantum limit scaling are derived asymptotically and cannot be expected to be literally the same for finite statistics.
We numerically predict the expected scaling for our parameter regime by
substituting $P_\mathrm{exp}=P_\mathrm{QM}\approx 0.85$ into Eq. \ref{eq:confidence}, and solving for $\eta$ as a function of $N$.
This is plotted as the solid blue line which has a slope of $-0.566$.
This agrees quite well with our experimentally measured scaling.
For the tripartite case, we use $20$ repetitions for $N$ from $1$ to $75$ (dark points in Fig. \ref{fig::infidelity}b inset), fitting to these data gives a scaling of $\eta\propto N^{-0.801}$ (dashed green line Fig. \ref{fig::infidelity}a inset), which significantly exceeds the standard quantum limit. 
The deviation from the ideal $N^{-1}$ Heisenberg-like scaling is mainly due to the imperfections in state preparation and the fact that we operate in the finite-statistics regime.
To account for this, we select a data range from from Fig. \ref{fig::infidelity}(b) where all the samples pass the nonlocal game {(from $N\in\left[10,100\right]$)} (grey points in the same inset) and fit the result (blue dashed line).
Doing so yields a scaling of $\eta \propto N^{-0.907}$.
We compare this to the numerically-predicted scaling behaviour for ideal GHZ states as before: we use Eq. \ref{eq:confidence} with $P_{exp}=1$ and obtain $\eta \propto N^{-0.935}$ (blue line).
This again agrees well with our experimentally measured scaling, with a small discrepancy coming from failed events before the displayed range.
Both our two- and three-photon data qualitatively show the expected trends.

\begin{figure}[t]
    \centering
    \includegraphics[width=\linewidth]{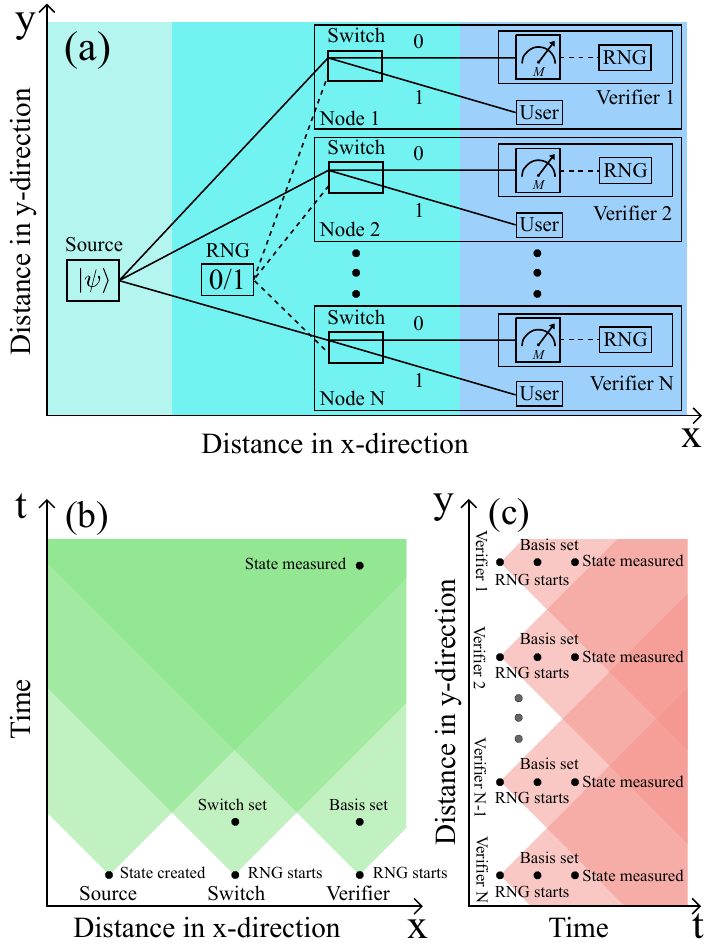}
    \caption{\textbf{Proposal for Device Independent Quantum State Certification.} a) A heralded N-photon source send photons to N space like separated nodes.  A random number generator (RNG) also distributes signals to the N nodes, which are used to synchronize trusted switches in each node. In addition to these switches, each node hosts a user, and verifier measurement station.  The verifier measurement stations operate device independently. Panels b) and c) show the space-time diagrams along the x and y directions, respectively.
    In the Methods Sec. C a loophole analysis of this setup is presented.}
    \label{fig::loophole}
\end{figure}

\section*{Discussion}
We have presented an experimentally feasible method for QSC.
To do so, we used active switches to deterministically implement a random sampling of multi-photon states from two- and three-photon sources.
A verifier measures the sampled states, issuing a statistically rigorous certificate to the user that the remaining states have {average} fidelity to a target state above some threshold.
From the user's point of view the only noticeable effect of the verifier is a slight reduction in the counting rate that does not scale with the number of photons in the entangled state.
Moreover, we have demonstrated two distinct scaling behaviors {related to different features of nonlocal games; i.e. the winning probability.}
While we have shown this technique for Bell states and GHZ states, it can be {generalized to} \textit{any}  states for which a robust self-testing protocol exists.
We also mention that although the theory underlying our work is DI, our implementation contains loopholes and is thus not device independent. However, to guide future DI implementations, Fig. 5 presents a potential extension of our experiment which could close the loopholes needed to claim device independence. See Methods Sec. C for a full description of this proposal. 

To summarize, we have performed quantum state certification (as opposed to verification), showing how to use active switches randomly-sample a subset of the available states to certify the remaining states. We have demonstrated the efficiency of quantum state certification, experimentally beating the standard limit of $N^{-0.5}$ with $N^{-0.801}$. Moreover, quantum state certification removes the assumption of identically-distributed states, and can be implemented in a device independent manner. While our experiment is not device independent, we have proposed how this could be achieved in future work.
Finally, the relative ease of our realization, requiring only local measurements and low postprocessing complexity, means that quantum state certification could be readily applicable to other quantum systems, such as trapped ions \cite{joshi2023exploring,zhang2017observation}, cold atoms \cite{shaw2024benchmarking,graham2022multi}, and superconducting circuits \cite{cao2023generation}. We thus anticipate that our work could be used to benchmark future large-scale quantum devices.

{After the completion of our experiment we became aware of closely related work \cite{martins2024experimental}.}


\section*{MATERIALS AND METHODS}
\subsection{Deterministic Random Sampling with Optical Switches}

As discussed in the main text, we use optical switches to redirect photons to the verifier station.
For both our two- and three-photon experiments we use fiber-coupled optical switches. For our two-photon experiment the swtiches are based on integrated electro-optic modulators (EOM), while our three-photon experiments use microelectromechanical system (MEMS) based switches.
For our two-photon work, we employ one Agiltron 2x2 Nanospeed switch and one BATi Inc. 2x2 Nanona fiber switch. These switches feature rise and fall times of approximately 100 ns and 50 ns, respectively. Both devices have a maximum repetition rate of 1 MHz and exhibit a cross-talk between output ports of around 20 dB.
The transmission of each switch is $\approx79\%$, resulting in a two-photon transmission of $0.79^2\approx0.62$.
In the three-photon experiment, we use three identical Agiltron 1x2 fiber optical MEMS switches, each with a rise and fall time of approximately 2 ms, a maximum repetition rate of 5 Hz, and a cross-talk between output ports ranging from 50 to 75 dB.
The transmission of each switch is $\approx95\%$, resulting in a three-photon transmission of $0.95^3\approx0.85$.
In both cases, we synchronize the optical switches and set them to alternate between the user and verifier stations with a nominal $50\%$ duty cycle.

To ensure that states randomly directed to the verifier, {there are two essential conditions need to be satisfied: (1) the verifier (or user) should receive the full copy. For example, for the three photon experiment this means that all three photons should be sent to the same measurement station. 
To this end, we synchronize the optical switches with an external trigger signal.
This is verified in Fig. \ref{fig::randomness}a and in the Supplementary Material Sec. C.
The discrepancy in Fig. \ref{fig::randomness}a between the ideal switching efficiency ($\epsilon = 1$) is due to the ``synchronization efficiency,'' discussed in the main text, we can be reduced by non-zero crosstalk between switch output ports. The EOM-based switches used in the two-photon experiment exhibit higher crosstalk than the MEMS-based switches, leading to a non-zero probability of single photons ending up in the ``wrong'' output port, thereby reducing $\epsilon$.
(2) Distribution of each copy is random.
Although we drive the switches with a deterministic signal, we utilize the intrinsic randomness} from the SPDC process, {wherein each photon pair is generated at a random time. 
Therefore when each state is generated it can randomly lie in a time slot which will distribute it to the verifier or user. 
However, this requires the $M$-photon generation rate to be substantially lower than the switching rate, to avoid having one of more copies of the state in each time slot. 
Consequently, we set our switches speeds to  800 kHz and 5 Hz repetition rates for the two-photon Bell states and three-photon GHZ states, respectively.
In both cases, this is significantly higher than the respective state generation rates of $\approx 62$ kHz and $\approx 0.5$ Hz. 
Fig. \ref{fig::randomness} verifies the validity of this approach.

\subsection{Measurement Strategy}

Our DI-QSC protocol is based on converting the CHSH and Mermin inequalities into two- and three-photon non-local games, respectively.
This works by considering the different observables used to violate the respective inequalities, and associating each observable with a binary classical input ascribed to each local party.
In particular, for the bipartite case we have the possible inputs $(i_1,i_2)=\{(0,0), (0,1), (1,0), (1,1)\}$, and for the tripartite case we have $(i_1,i_2,i_3)=\{(0,0,1), (0,1,0), (1,0,0), (1,1,1)\}$.
These inputs then refer to the following observables
\begin{equation}
\begin{split}
\mathcal{M_{\text{CHSH}}} =& \Bigl\{X_1 \otimes \frac{X_2 + Z_2}{\sqrt{2}}, X_1\otimes \frac{X_2 - Z_2}{\sqrt{2}},\\
& Z_1\otimes \frac{X_2 + Z_2}{\sqrt{2}}, Z_1 \otimes \frac{X_2 - Z_2}{\sqrt{2}} \Bigr\}, \\
\mathcal{M_{\text{Mermin}}} =& \bigl\{Y_1 \otimes Y_2\otimes X_3, Y_1\otimes X_2\otimes Y_3,\\
& X_1\otimes Y_2\otimes Y_3, X_1\otimes X_2\otimes X_3\bigr\}.
\end{split}
\label{eq:measurements}
\end{equation}
In both cases, the result is 4 different multipartite measurements that must be made randomly.

To implement the random selection of measurements for our two-photon experiment, we use a 50/50 beamsplitter in the path of each photon to direct that photon to physically different measurement apparati corresponding to the binary inputs (Fig. \ref{fig:setup}b). This enables us to randomly select our measurements at very high counting rates, but it comes at the cost of opening the freedom-of-choice loophole. Nevertheless, since our measurement stations are not space-like separated our experiment is open to the locality loophole. Thus, since our experiment is anyways open to loopholes, we chose the measurement bases passively for simplicity.
Since our three-photon rate is significantly lower, we instead only have a measurement device per photon. {The random classical inputs are then generated by a commercial quantum random number generator (QRNG), the Quantis QRNG USB from ID Quantique. These outputs of this QRNG are set to trigger motorized waveplates that change the measurement settings.}
To further ensure that our sampling is independent, {we switch the settings every second (which is faster than the generation rate) and only analyze measurements time slots that contain no more than one three-photon event.}

\subsection{Loophole Analysis}

\textbf{Standard Loopholes.} Device-independent protocol implementations ensure robust protection against any external manipulation or influence on the protocol, maintaining reliable and trustworthy outcomes. 
In principle, even if the entire apparatus were controlled by an untrusted party, the protocol would still deliver correct results. 
Implementing DI-QSC requires a loophole-free violation of a Bell inequality (the CHSH or Mermin inequality in our case), and thus the common loopholes---\textit{i.e.} the (1) locality, (2) freedom of choice, and (3) detection---must be closed \cite{Larsson_2014}.
We do not achieve this in our work.
While loophole-free Bell tests have recently been conducted {\cite{Hensen2015Loophole,Shalm2015strong,Giustina2015Significant}}, such demonstrations are rare due to the stringent infrastructure requirements involved.
Nevertheless, in Fig. \ref{fig::loophole}a we sketch an N-partite extension of our experiment that could address these three loopholes, as well additional ones related to multi-partite QSC.
In order to close the locality loophole (1), a source at a central location generates the N entangled photons which are distributed to N space-like separated nodes. 
The measurement basis must then be set after the state are generated, which can, in principle, be easily enforced (Fig. \ref{fig::loophole}b and c).
To address the freedom-of-choice loophole (2) the measurement basis must be randomly set.
In our three-photon experiment, we satisfy this by implementing our basis choice using a quantum random number generator.
In our two-photon experiment, the basis is set randomly using 50:50 beamsplitters. 
While this is a random process, it opens the loophole since the photons could carry hidden variables determining whether they are transmitted or reflected to one measurement setting or the other.
Finally, to ensure that the detection loophole (3) is closed the overall transmission between the source and the detectors (including the detection efficiency) must remain over a certain threshold. This is $66.7\%$ and $75\%$ for the two and three-photon experiments, respectively {\color{blue}\cite{2008PhRvL_Cabello}}.

\textbf{Preparation post-selection loophole.} While the above loopholes apply to all Bell tests, current methods of creating multi-partite entangled states open up an additional loophole, which we refer to as (4) the preparation post-selection loophole.  
This loophole is one of the primary challenges for a loophole-free mutli-partite demonstration of quantum nonlocality, which has not yet been realized \cite{Huang2022Experimental}. Closing this loophole is required for DI-QSC of multi-photon states.
The preparation post-selection loophole comes from the fact that we use non-heralded gates to build three-photon entanglement from two-photon entangled states.
We can only be sure that we have successfully generated the entangled state when one photon is detected in each output port, see Ref. \cite{Huang2022Experimental} for a discussion of the implications.
In order to close the post-selection loophole, one could use heralded sources of multi-photon entanglement, as in Ref. \cite{cao2024photonic}, which can be achieved with either deterministic sources or non-deterministic SPDC as we use here. 
Unfortunately, the heralding efficiency in such experiments is still relatively low, and thus opens the detection loophole.

\textbf{Loopholes Specific to DI-QSC.} Thus for we have discussed loopholes only related to loophole free violations of Bell inequalities, which must also be closed for DI-QSC. 
However, DI-QSC brings additional loopholes.
Here we identify and discuss two additional loopholes, (5) the switching-loss loophole and (6) the random sampling loophole.
Note that we also assume that the switches belong to the measurement nodes, and are thus trusted devices.
We will now discuss how our proposal in Fig. \ref{fig::loophole} addresses these loopholes.

We already alluded to the switching-loss loophole in the main text. The intuition is relatively straight-forward: upon insertion of the switches used to implement the random sampling, the transmission of the N-photon state from the source to the verifiers should remain above the threshold needed to close the detection efficiency loophole.
There are two contributions that reduce this transmission.
The first is simply the transmission loss of the switches. 
In our case, for the two-photon experiment the transmission is $\approx0.79$ per switch, while for the three-photon experiment it is $\approx0.95$ per switch.
This results in an overall two-photon transmission of $0.79^2=0.62$, and a three-photon transmission of $0.95^3=0.85$.
Thus while our two-photon switches are too lossy, our three-photon switches could close it.

The other source of loss is the synchronization efficiency, already discussed in the main text and in the Methods Sec. A.
For completeness we discuss it here in the context of loopholes.
When an N-photon state is selected to be sent to the verifier, all N photons must simultaneous be routed.
If some photons are sent to the user and others in the verifier, these events introduce loss.
If this loss reduces the source-to-verifier transmission too much the detection loophole is opened.
This is the reason that we use synchronized active switches.
If one uses ``passive switches'' (i.e. beamsplitters) this loss scales to zero exponentially fast with the number of photons in the entangled state. 
We call this the passive synchronization efficiency, and plot it for the case of 50:50 beamsplitters in Fig. \ref{fig::randomness}a.
Using synchronized active switches, on the other hand, can in principle route the photons without introducing any additional loss.
Experimentally we achieve synchronization efficiencies of $0.9439 \pm 0.0041\%$ and $0.9997 \pm 0.0066\%$ for our two- and three-photon experiments, respectively.
Which could be sufficiently high to remain above the detection efficiency threshold if other experimental losses were low enough.

The final loophole we discuss is the random sampling loophole (6).
If the sampling process is not genuinely random, the validity of the protocol is compromised.
The challenge in enforcing this experimentally is that the switches, which are located in space-like separated nodes, must be synchronized and the decision of whether to route the N-photon state to the verifier or to the user must be made after the source produces a state, just as is done in choosing the basis to close the standard freedom-of-choice loophole.
Nevertheless, this can be enforced with space-like separation.

\textbf{A Loophole Free Proposal.}
To show how our proposed setup can achieve close these 6 loopholes we will now analyze each component shown in Fig. \ref{fig::loophole}a, moving from left to right. Corresponding spacetime diagrams for the $x$- and $y$-directions are presented in Fig. \ref{fig::loophole}b and \ref{fig::loophole}c, respectively. 
We first assume an ideal photon source capable of generating a heralded multi-photon state, \textit{i.e.} one that closes loophole 4.

The next element in Fig. \ref{fig::loophole} is a random number generator (RNG), which produces a random binary output (‘0’ or ‘1’) to determine whether the optical switches will direct the state to the verifier or the user. 
It is essential that the RNG operates in a manner such that it selects the user or verifier before the photon source can signal to it (closing the loophole 6), and it can signal to the switches before each of the N photons can arrive at their respective node (so the switches can be synchronized and close loophole 5). 
Note that in our experiment we use randomness in the photon generation time to implement the random sampling, which, although random, does not close this loophole.
Achieving this requires aligning the space-like separation with the speed and timing of the electronics involved. 
In may also be possible to close this loophole by making the appropriate shielding assumptions as done in recent DI experiments \cite{liu2021device,liu2022toward}.

Once the switches are configured, two scenarios arise. If the state is routed to the user, no loopholes need to be addressed (unless the user also wishes to operate device independently, but we will not focus on that here). 
Conversely, if the states are directed to the verifier, device independence is necessary; \textit{i.e.} loopholes 1-3 must be closed.
Doing so requires the $N$ verifiers to select their measurements randomly independently and without correlation to the photon source. 
To ensure freedom of choice, the measurement settings must be determined before any signal from the source can influence a verifier. 
In photonics, this is typically achieved using a random number generator (RNG) in combination with a fast Pockels cell.

Additionally, all $N$ verifiers must be space-like separated from each other to prevent any {causal influence} in their measurements. 
This requirement is represented in the spacetime diagram for the $y$-direction in Fig. \ref{fig::loophole}c. 
Finally, addressing the detection loophole necessitates achieving a system detection efficiency above a specific threshold. 
To close this loophole, the setup must employ low-loss optical components, high-efficiency detectors, and sufficiently high coupling efficiencies between free space and fiber.
Overall implementing DI-QSC is a challenging goal, but no more challenging than implementing most other multi-partite DI protocols.
\\
\\

\noindent\textbf{Acknowledgements.} 
We are grateful to Wenhao Zhang for insightful discussions and comments on the manuscript, and to Alessandro Trenti and Martin Achleitner for randomness discussions.
This research was funded in whole, or in part, by the European Union (ERC, GRAVITES, no. 101071779) and its Horizon 2020 and Horizon Europe Research and Innovation Programme under grant agreement no. 899368 (EPIQUS) and no. 101135288 (EPIQUE) and the Marie Skłodowska-Curie grant agreement no. 956071 (AppQInfo). Views and opinions expressed are however those of the author(s) only and do not necessarily reflect those of the European Union or the European Research Council Executive Agency. Neither the European Union nor the granting authority can be held responsible for them. Further funding was received from the Austrian Science Fund (FWF) through 10.55776/COE1 (Quantum Science Austria), through 10.55776/F71 (BeyondC) and 10.55776/FG5 (Research Group 5) and from the Air Force Office of Scientific Research under award number FA9550-21-1-0355 (QTRUST) and FA8655-23-1-7063 (TIQI); the financial support by the Austrian Federal Ministry of Labour and Economy, the National Foundation for Research, Technology and Development and the Christian Doppler Research Association is gratefully acknowledged. L.A.R. acknowledges support from the Erwin Schrödinger Center for Quantum Science and Technology (ESQ Discovery).
For the purpose of open access, the author has applied a CC BY public copyright license to any Author Accepted Manuscript version arising from this submission.
\\

\bibliography{QSC.bib}

\clearpage
\newpage
\hypertarget{sec:appendix}
\onecolumngrid 
\appendix
\newpage
    \renewcommand{\thesubsection}{\Alph{subsection}}
    \setcounter{equation}{0}
    \numberwithin{equation}{subsection}
    \setcounter{figure}{0}
    \renewcommand{\thefigure}{A.\arabic{figure}}

\section*{Supplementary Material}
\subsection{Two-Photon Source}
\label{sec:details_setup}

\noindent For our two-photon experiment we use a type-0 spontaneous parametric down-conversion (SPDC) source as in Ref. \cite{type0_source_Ursin}, illustrated in Fig. \ref{fig:setup_2photon}. The source uses a periodically-poled Lithium Niobate (ppLN) crystal that is placed in a Sagnac loop and pumped by a narrowband continuous-wave laser at 775.06 nm.
Before the Sagnac loop, a half waveplate (HWP) sets the pump polarization to $45$°. After the entangled photon pairs are created, they pass through a long pass filter (LPF) in order to suppress the remaining pump light.
The entangled photon pairs are then seperated with a $100$ GHz dense wavelength division multiplexing (DWDM) module. Due to energy conservation in the SPDC process, entangled photon pairs are separated in DWDM channels symmetric around the center of the SPDC emission spectrum, which is at $1550.12$ nm.
In this experiment we use entangled pairs from DWDM channels $32$ and $36$, with center wavelength $1551.72$ nm and $1548.51$ nm, respectively.
Each photon is sent to an individual optical switch (OS), which are synchronized with each other to route photon pairs either to the verifier or the user. At the verifier, each photon passes a 50:50 BS to ensure randomness in the choice of the measurement basis. Depending on if the photon is transmitted or reflected, a different measurement is performed. At the user side, the measurement settings can be set to an arbitrary basis, using waveplates.

\begin{figure*}[h]
\centering
\includegraphics[width=0.8\textwidth]{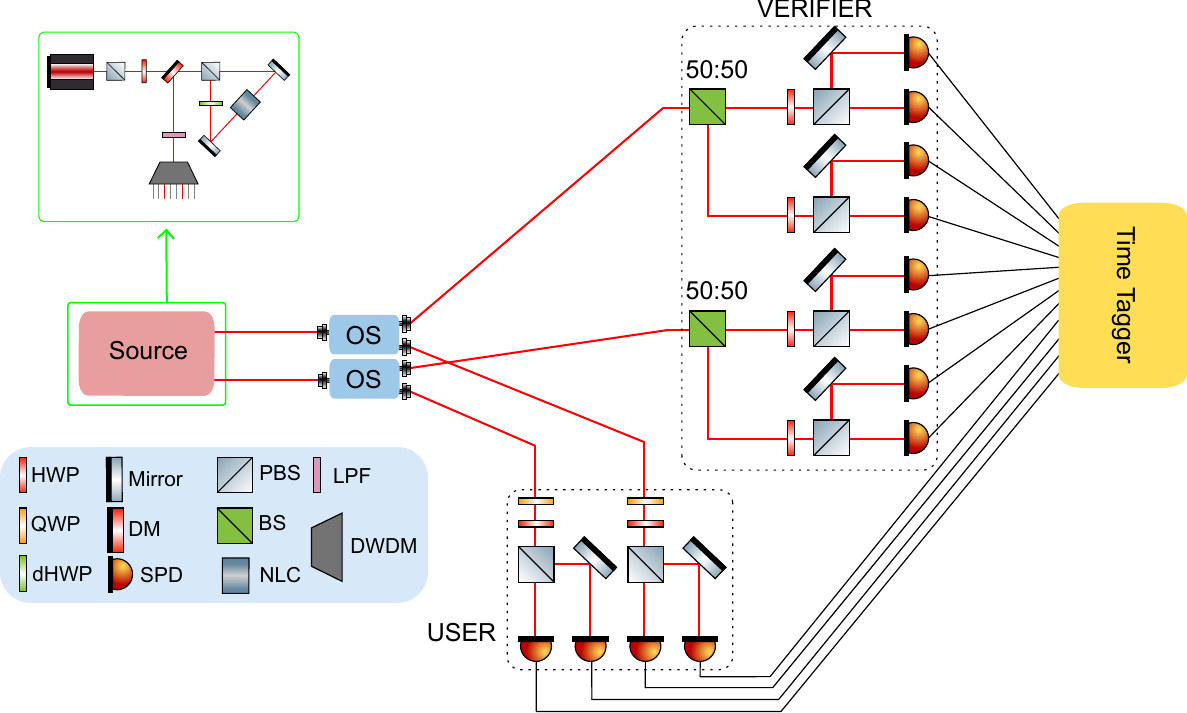}
	\caption{\textbf{Two-Photon Experimental Setup.} A continuous-wave laser at 775.06 nm pumps the nonlinear crystal (NLC) to create entangled photon pairs in the telecom C-band centered around 1550.12 nm via type-0 SPDC (spontaneous parametric down-conversion). The DWDM divides the total photon spectrum into several channels. The entangled photons are found in channels equidistant from the center of the spectrum. Each photon is forwarded to one of the two optical switches that are driven by synchronized square signals with 50\% duty cycle. This means that approximately half of the photon pairs are sent to the verifier and the other half to the user. The photons sent to the verifier are split at a 50:50 BS, in order to randomly choose the measurement settings. Coincidence detection is performed using a time tagger. The photons at the user side can be used for some other purpose. In our experiment, they are measured with waveplates and polarizing beamsplitters. Abbreviations: HWP, half-wave plate; QWP, quarter-wave plate; dQWP, dual-wavelength quarter-wave plate; DM, dichroic mirror; SPD, single photon detector; PBS, polarizing beam splitter; BS, beam splitter; LPF, longpass filter; DWDM, dense wavelength division multiplexer; OS, optical switch;}
\label{fig:setup_2photon}
\end{figure*}

\subsection{Three-Photon Source}
\begin{figure*}[h]
\centering
\includegraphics[width=0.8\textwidth]{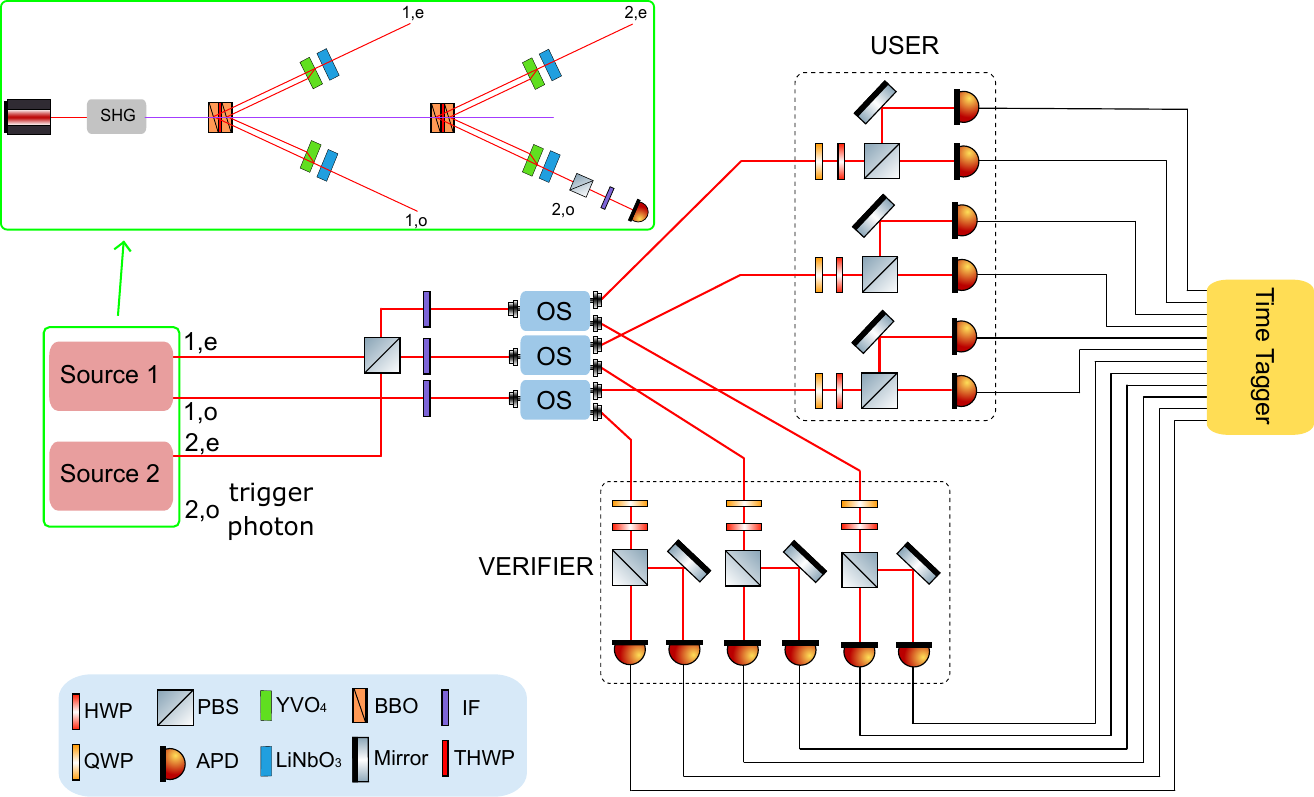}
	\caption{\textbf{Three-Photon Experimental Setup.} The two sources are pumped with a pulsed beam at a wavelength of 390 nm, produced from second-harmonic generation (SHG) of a Ti:Sapph laser. The photons generated from the SPDC process in the BBO crystals therefore exhibit a wavelength of 780 nm. One photon of each created entangled photon pair (1,e and 2,e) is further sent to a PBS to create a three-photon entangled state (Hong-Ou-Mandel interference). Each of the three photons (1,e, 2,e and 1,o) enters one optical switch, respectively, that are driven by a square wave at 5 Hz with 50 \% duty cycle. Therefore, as in the two-photon setup, half of the photons are sent to the verifier side and the other half to the user side. The measurement settings at the verifier side are randomly chosen by a quantum random number generator (QRNG). The photons are detected by APDs and their coincidences are recorded by a time tagger. The fourth photon (2,o) is triggered in horizontal (H) polarization via another PBS and serves as a trigger photon. Again, the user can perform any arbitrary measurement. Abbreviations: BBO, Beta Barium Borate; THWP, true zero-order half waveplate; YVO$_{4}$, Yttrium orthovanadate; LiNbO$_{3}$, Lithium niobate; OS, optical switch; HWP, half-wave plate; QWP, quarter-wave plate; PBS, polarizing beam splitter; APD, Avalanche Photodiodes; IF, interference filter;}
\label{fig:setup_3photon}
\end{figure*}
\noindent A sketch of our three-photon experiment is shown in Fig. \ref{fig:setup_3photon}. It consists of two sandwich-like entanglement sources \cite{zhang2015}. Each sandwich-like entanglement source consists of two type-II BBO crystals with a true-zero-order half-wave plate (THWP) in between. When pumped by a laser, the first BBO produces two photons in the state $\ket{H}_e\ket{V}_o$, where the indices e and o denote an (extra-) ordinary photon. The THWP further rotates this state into $\ket{V}_e\ket{H}_o$, while the pumping beam remains unchanged and is used to pump the second BBO, that also creates a photon pair in the state $\ket{H}_e\ket{V}_o$. After applying temporal and spatial compensations, the two possible ways of generating twin photons become indistinguishable.
Hence, the entangled state $\frac{1}{\sqrt{2}}(\ket{H}_e\ket{V}_o - \ket{V}_e\ket{H}_o)$ is generated. In our case, we use two such sandwich-like entanglement sources. 
in order to generate a multipartite GHZ state, we transform the state described above into $\frac{1}{\sqrt{2}}(\ket{HH} + \ket{VV})$ in the first source. We then take a single photon, prepared in the state $\frac{1}{\sqrt{2}}(\ket{H} + \ket{V})$, from the second source. We do so by triggering one photon of the pair into horizontal polarization state and then applying local unitary to other heralded photon. 
By overlapping this heralded single photon with the entangled two-photon state at a polarizing beam splitter (PBS), the three-photon state $\frac{1}{\sqrt{2}}(\ket{HHH} + \ket{VVV})$ is created via Hong-Ou-Mandel (HOM) interference. A filter with a bandwidth of 3 nm is inserted in each photon's path. After the photons enter the optical switches (OS) they are randomly sent to the verifier or the user side. Each side consists of an arrangement of quarter- and half-wave plates, PBSs and Avalanche photodiodes (APD) to analyze the polarization of the photons and to measure in any arbitrary basis.
In the verifier's side, the waveplates are rotated to one of two measurements based on the output of a commercial quantum random number generator (QRNG) from ID quantique. 

\subsection{Synchronization of the optical switches}
\noindent In order to synchronize the optical switches, we have to ensure that we modulate the OSs in phase. To achieve this, the outputs of the function generator producing the square signals driving the switches are synchronized. Synchronization can be achieved by observing both the wanted coincidence counts (those between detectors only in the verifier or the user side) and the unwanted cross coincidence counts (those occurring between detectors in the user and verifier's sides). By adjusting the phase between the square signals accordingly, the desired coincidence counts are maximized, while the unwanted cross events are minimized. The corresponding resulting coincidence events are shown in Fig. \ref{fig:cross_cnts}. Note that here all possible cross coincidences have been summed.

\begin{figure}[h]
\centering
\begin{minipage}{.5\textwidth}
  \centering
  \includegraphics[width=1\linewidth]{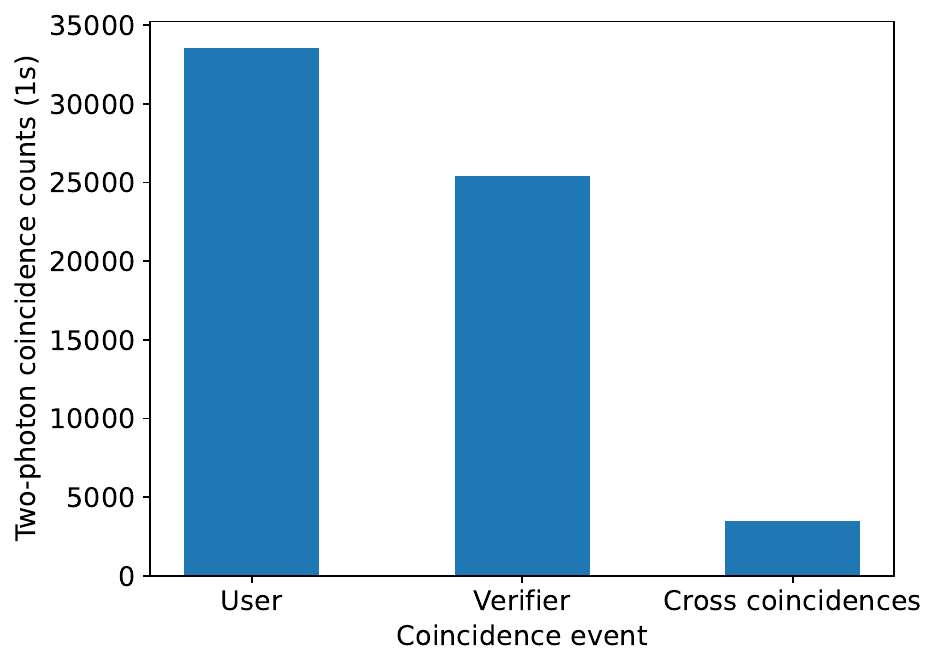}
\end{minipage}%
\begin{minipage}{.5\textwidth}
  \centering
  \includegraphics[width=1\linewidth]{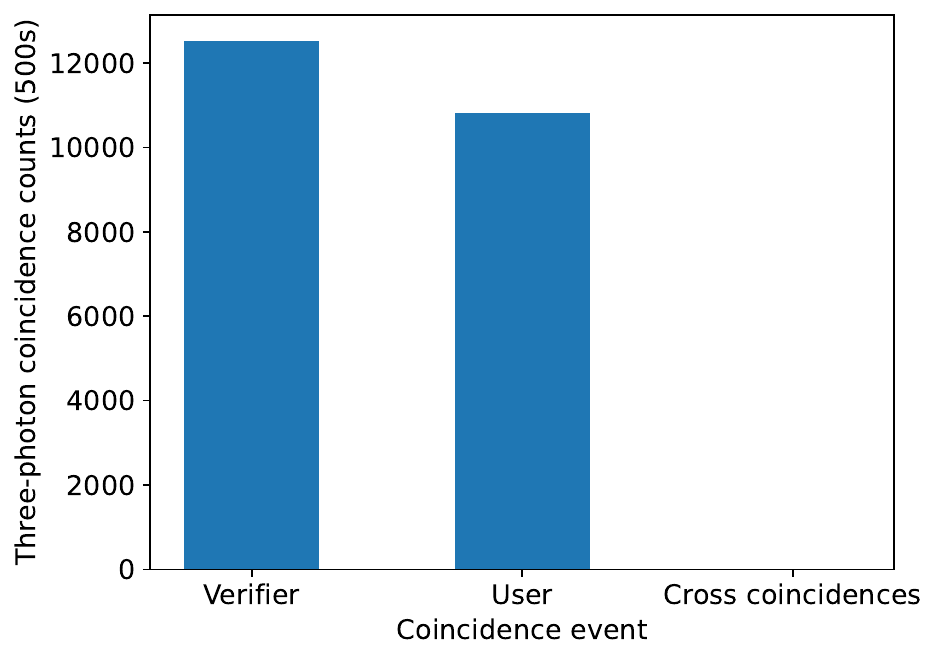}
\end{minipage}
\caption{\textbf{Synchronization of the optical switches.} Coincidence counts between the different detectors for the two-photon setup (left) and three-photon setup (right). Coincidences were detected between channels on the user side, on the verifier side and cross coincidences are those between both sides.}
\label{fig:cross_cnts}
\end{figure}

\subsection{Randomness quantification}

\noindent 
The assumption that each state is randomly distributed between verifier and user is crucial for quantum state certification. 
In order to provide evidence that a given sequence of events is truly random, there exists a series of random number generation tests according to the National Institute of Standards and Technology (NIST) \cite{bassham2010}. Using this tool, a sequence of binary bits is said to be randomly generated with a certain confidence if it passes all the tests.

In our case, the random distribution between the user and verifier results from the inherent randomness in the SPDC process, which we take advantage of by setting the OSs to switch faster than the multiphoton events occur. 
However, the distribution of both the two-photon Bell state and the three-photon GHZ state between verifier and user is slightly biased, primarily due to different detection efficiencies of the SPDs and different insertion losses between of the OSs. 
To be precise, for the two-photon Bell state, the bias is around 43 \% for the verifier and 57 \% for the user. 
For the three-photon GHZ state, the bias is around 53 \% for the verifier and 47 \% for the user. Unfortunately, the NIST tests do not tolerate any bias. Therefore, we have to apply a randomness extractor that removes the bias but keeps the intrinsic randomness of the data. For this reason,
we calculate the conditional min-entropy $H_{cond}$ of a random variable $A$ given $B$, which is a measure of the worst-case unpredictability of $A$ given knowledge of $B$. In other words, it tells us how much uncertainty remains about $A$ when $B$ is known \cite{Konig_2009}. 
$H_{cond}$ accounts for scenarios where side information could reduce the effective entropy of the data. This ensures the randomness we extract is robust, even in the presence of potential correlations or leaks. It can be seen as a measurement of the effectiveness of the strategy to guess the most likely output. A high $H_{cond}$ (close to 1) means that the output of our source is unpredictable, whereas a low $H_{cond}$ (close to 0) means that the output is predictable. $H_{cond}$ is calculated based on the maximum relative entropy $D_{max}$ of some bipartite density operator $\rho_{AB}$, as shown in equation \ref{eq:cond_min_entropy}.
The min-entropy $H_{min}$ is calculated based on the more likely output $p_i$, with $i$ = \{0, 1\}, as shown in equation \ref{eq:min_entropy}.


\begin{equation}
    H_{cond} = -\text{inf $D_{max}$}(p_{AB} || I_A \otimes \sigma_B)
    \label{eq:cond_min_entropy}
\end{equation}

\begin{equation}
    D_{max} = \text{inf}\{\lambda:p\leq2^{\lambda}\sigma\}
    \label{eq:dmax}
\end{equation}

From $H_{cond}$ we can infer the amount of random data we can extract from our biased data. We are using Toeplitz hashing \cite{krawczyk1995}, as it allows us to choose the amount of random data we want to extract. The same procedure has already been used in previous work for extracting randomness from weak random sources, such as in \cite{Ng_2023}. For Toeplitz hashing we need a suitable Toeplitz matrix that is multiplied by a vector containing our raw random data. The Toeplitz matrix can be generated using any reliable reference random data. The vector resulting from the matrix multiplication then contains the extracted random data to which we apply the NIST randomness tests. Before applying the tests, we assume the null hypothesis that the data are random, which we wish to confirm. The tests are statistical and each one yields a p-value to demonstrate the strength we confirm or reject the null hypothesis with. In this case, the higher the p-value the better the randomness. To analyze the data we first set a certain significance level $\alpha$. If p-value $\geq \alpha$, the test is passed and the null hypothesis is confirmed. We set $\alpha$ into 0.01, meaning that we can confirm randomness with 99 \% confidence. To generate a random bit sequence from our experiment, we assign a multiphoton coincidence event on the verifier side to ``1'' and a multiphoton coincidence event on the user side to ``0''. From the resulting data, we calculate a conditional min-entropy of $H_{cond, Bell}$ = 0.8091 for the two-photon case and $H_{cond, GHZ}$ = 0.9140 for the three-photon case. This value is equal to the percentage of randomness we can extract from our raw data via Toeplitz hashing. To generate the Toeplitz matrix, we use our QRNG. The vectors resulting from the matrix multiplication are used to perform the NIST tests. We recalculate the conditional min-entropy for the extracted random data to show how much it has changed close to an ideal value. We calculate an improved min-entropy of $H_{cond, Bell}$ = 0.9974 for the two-photon case and $H_{cond, GHZ}$ = 0.9973 for the three-photon case.\\
Additionally, we perform the randomness extraction with the von Neumann extractor \cite{peres1992}. In contrast to the Toeplitz extractor, we cannot set the amount of randomness we wish to extract here, but merely remove the bias from our raw data. This is done by assigning '0' to two subsequent bits '01' and '1' to two subsequent bits '10'. This procedure effectively removes the bias and still keeps the intrinsic randomness. However, one drawback of this method is that it diminishes our datasets by more than half. The amount data we collected for the three-photon GHZ state after having applied von Neumann extraction is not enough to perform the ninth test of the NIST test suite 'Universal Statistical Test'. For this reason, we show the NIST test results received from the Toeplitz extraction also in the main text, while showing the results from the von Neumann extraction only in this section, as an additional comparison (see fig. \ref{fig:nist_toeplitz} and \ref{fig:nist_von_neumann}).
\begin{figure}[h]
    \centering
    \begin{minipage}[t]{0.485\linewidth}
        \centering
        \includegraphics[width=\linewidth]{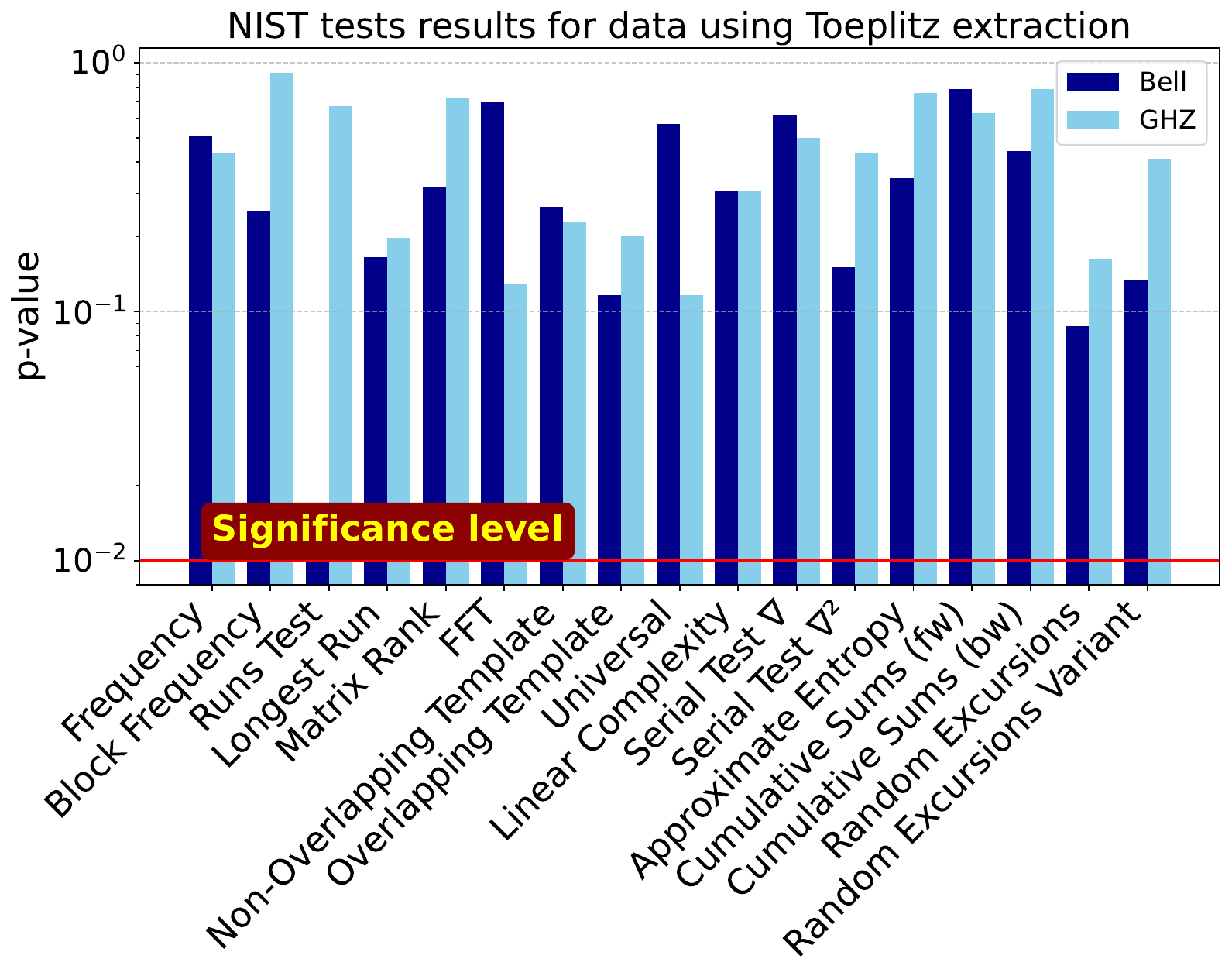}
        \caption{NIST test results after applying Toeplitz hashing to our raw data. The bars in dark blue correspond to the two-photon Bell states and the bars in light blue to the three-photon GHZ states.}
       \label{fig:nist_toeplitz}
    \end{minipage}
    \hfill
    \begin{minipage}[t]{0.485\linewidth}
        \centering
        \includegraphics[width=\linewidth]{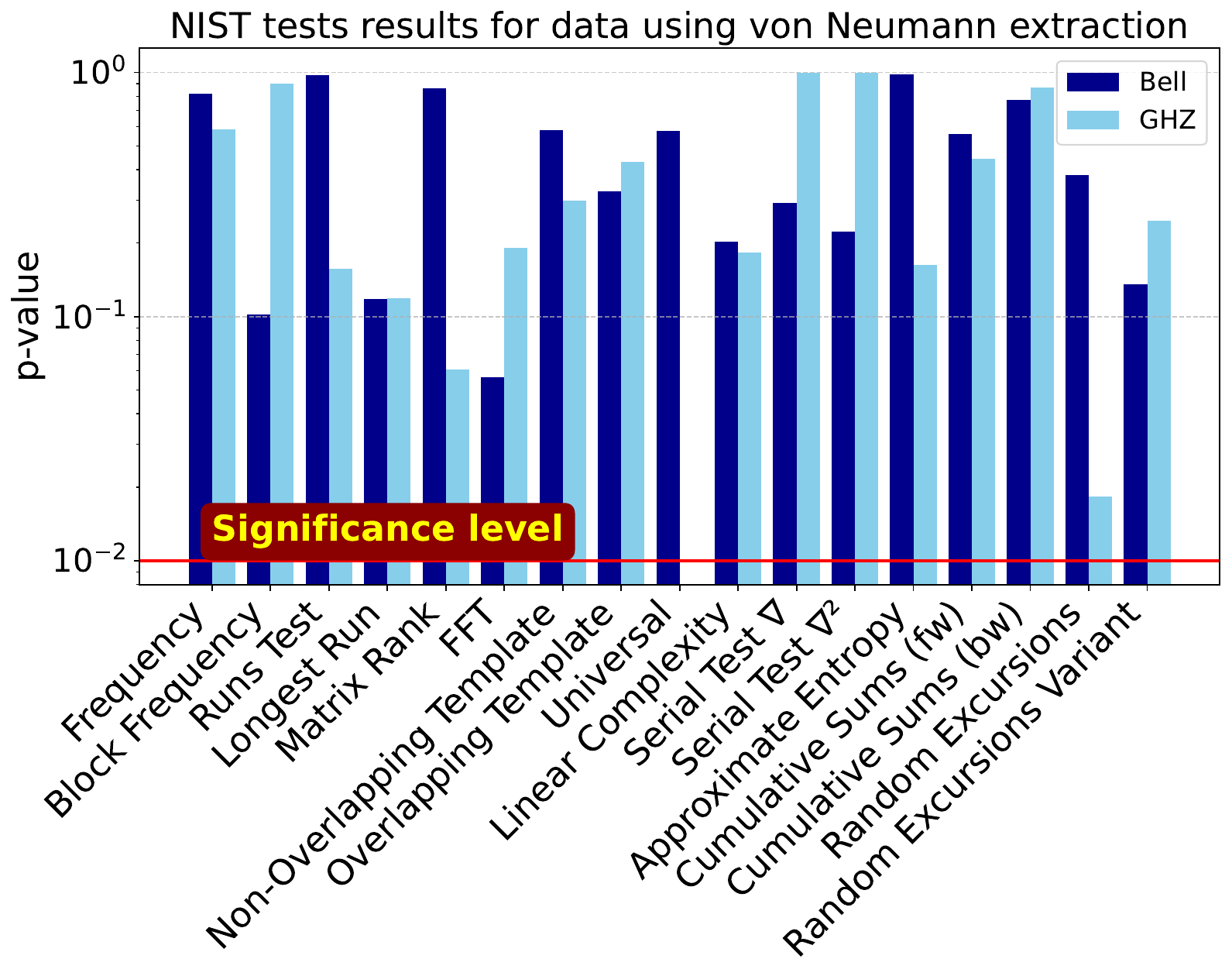}
        \caption{NIST test results after applying von Neumann randomness extraction to our raw data. The bars in dark blue correspond to the two-photon Bell states and the bars in light blue to the three-photon GHZ states. In the case of the von Neumann extraction, the data for the GHZ states are too strongly reduced to apply the “Universal Statistical Test”.}
      \label{fig:nist_von_neumann}
    \end{minipage}
\end{figure}

\subsection{Success probability}
To determine the overall probability of success, we calculate the ratio between the correct results and the number of samples used so far. For each sample, we perform a measurement that is randomly selected from the four possible settings. The outcome is either correct or incorrect, resulting in the curves shown in Fig. \ref{fig:succ_prob}. Each time the success probability decreases, an incorrect result has been detected. However, after enough rounds the curve asymptotically approaches $P_{exp}$, which is $P_{exp,Bell}$ $\approx$ 0.847 for the two-photon Bell state (Fig. \ref{fig:succ_prob}, top) and $P_{exp,GHZ}$ $\approx$ 0.977 for the three-photon GHZ state (Fig. \ref{fig:succ_prob}, bottom). The insets in Fig. \ref{fig:succ_prob} show the extended measurement of the success probability.
\begin{figure*}[h]
\centering
\includegraphics[width=0.8\textwidth]{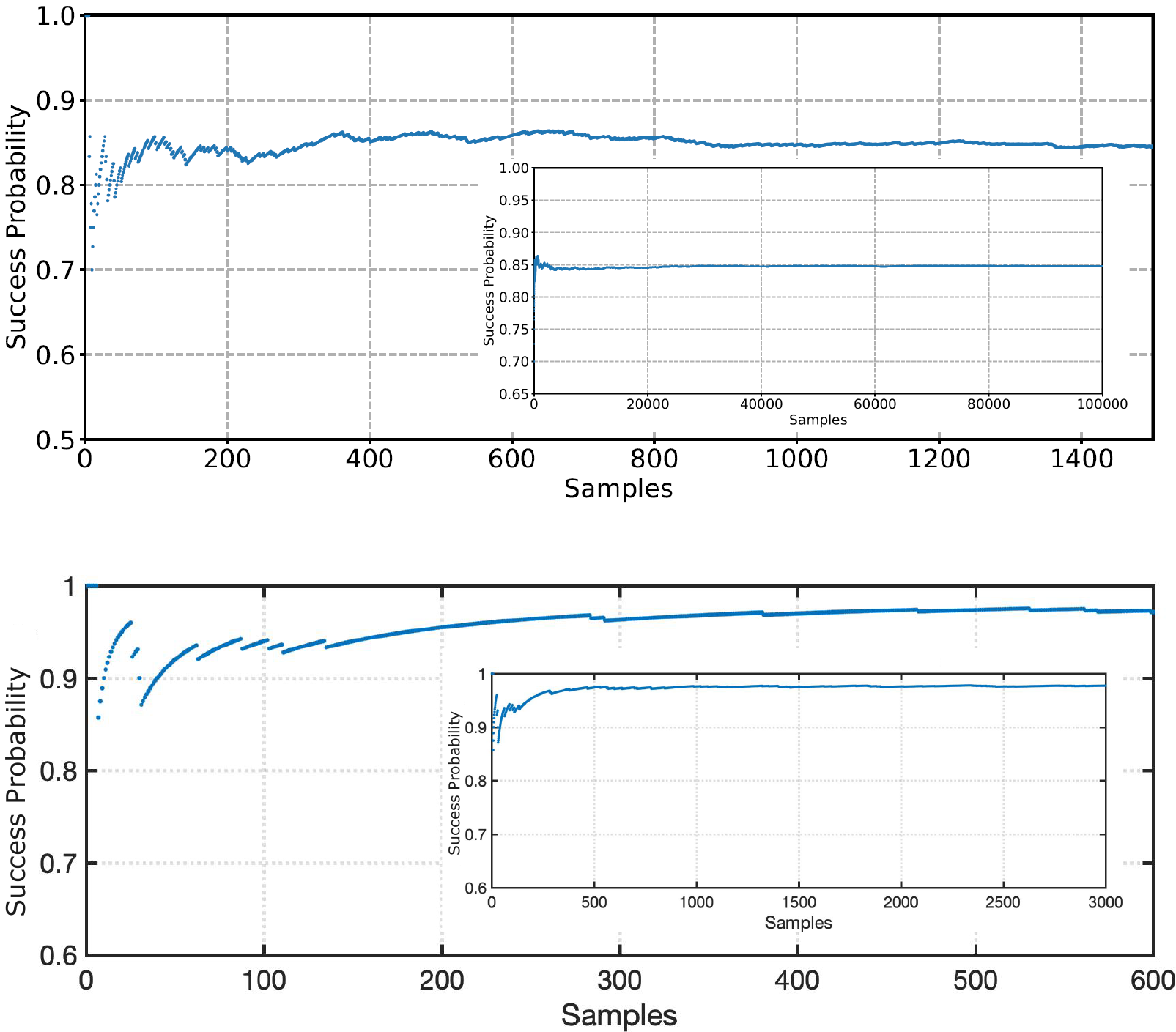}
	\caption{\textbf{Probability of success vs number of samples.} The upper panel shows the success probability in dependence of the number of samples for the two-photon Bell state, while the lower panel shows the same for the three-photon GHZ state. The insets in both figures show the extended measurements.}
\label{fig:succ_prob}
\end{figure*}

\subsection{Fidelity estimation by Witness and self-testing}
\textit{Device-dependent evaluation of fidelity by a witness}.-- We can determine the fidelity of the GHZ state by projecting it into the ideal GHZ state. The projective operator can be decomposed into local measurements, leading to a witness in the form of \cite{huang2011experimental,guhne2007toolbox}
\begin{align}
    \mathcal{W}_{GHZ}&=|GHZ_3\rangle\langle GHZ_3|= \frac{1}{2}A_3+\frac{1}{6}\sum\limits_{k=1}^3 (-1)^k M_k \\
    A_3&=(|H\rangle\langle H|)^{\otimes 3}+ (|V\rangle\langle V|)^{\otimes 3} \\
    M_k&=\left[\text{cos}(\frac{k\pi}{3})\sigma_x+\text{sin}(\frac{k\pi}{3})\sigma_y\right]^{\otimes 3},\ k=0,1,2
\end{align}
The fidelity of a physical state $\rho$ on the user side can be estimated by the evaluation of witness $F=\langle GHZ_3|\rho|GHZ_3\rangle=\langle W_{GHZ} \rangle=0.9679\pm0.0052$.

\textit{Device-independent valuation of fidelity by self-testing}.--{In a Bell inequality, entangled states can reach a maximal the value, called the quantum bound $\beta_Q$, which exceeds the classical bound $\beta_C<\beta_Q$ admitted by local hidden variable model. Experimentally, one typically observes a Bell value $\langle\mathcal{B}\rangle$ lower than the quantum bound, $\langle\mathcal{B}\rangle=\beta_Q-\epsilon$.
In this case, robust self-testing can be used to provide a lower bound on the fidelity $\mathcal{F}=1-f(\epsilon)$ between the physical state and a target state up to local isometries. Endeavors have been made to optimize the function $f$ to improve the robustness performance. 
Here we take use of the robust self-testing strategy of Ref. \cite{kaniewski2016analytic,zhang2018experimentally}. } In the bipartite case, the underlying inequality for self-testing is Clauser-Horne-Shimony-Holt (CHSH) inequality
\begin{equation}
    \mathcal{B}_{bell}=\langle A_0 B_0 \rangle+\langle A_0 B_1 \rangle+\langle A_1 B_0 \rangle-\langle A_1 B_1 \rangle<2
\end{equation}
where $A_0=X, A_1=Z$ and $B_0=\frac{X+Z}{\sqrt{2}}, B_1=\frac{X-Z}{\sqrt{2}}$.
{The corresponding relation between the DI-fidelity lower bound and the observed Bell violation is given by
\begin{equation}
    \mathcal{F}_{bell}\leq \frac{1}{2}+\frac{1}{2}\frac{\langle\mathcal{B}_{bell}\rangle - \beta^{*}}{2\sqrt{2}-\beta^{*}}, \quad \beta^{*}=\frac{16+14\sqrt{2}}{17}\approx 2.11
\end{equation}
In tripartite case, we apply the Mermin inequality
\begin{equation}
    \mathcal{B}_{ghz}=\langle A_0 B_0 C_0\rangle+\langle A_0 B_1 C_1 \rangle+\langle A_1 B_0 C_1 \rangle-\langle A_1 B_1 C_0 \rangle<2
\end{equation}
where $A_0=B_0=C_0=X$ and $A_1=B_1=-C_1=-Y$. Then the relation between this observed violation and the DI-fidelity is
\begin{equation}
    \mathcal{F}_{ghz}= \frac{1}{2}+\frac{1}{2}\frac{\langle\mathcal{B}_{ghz}\rangle - \gamma^{*}}{4-\gamma^{*}}, \quad \gamma^{*}=2\sqrt{2}.
\end{equation}
}

\subsection{Confidence Approximation}
The confidence level $1-\delta$ provided by the original QSC protocol is $1-\delta=1-\left[1-\mu+\mu e^{-D}\right]^N$, where the $D=D(P_\mathrm{exp}\parallel P_\mathrm{\eta})$ and all other notations are defined in the main text. In our work, we instead use an approximation, $1-\delta=1-e^{-D\mu N}$, to facilitate experimental implementation under realistic conditions. This approximation is convenient for practical use while maintaining sufficient accuracy. 

We demonstrate that the two formulas are equivalent when higher-order terms of their Taylor expansions are omitted. The original definition can be approximated as:
The original definition can be expressed as
\begin{align}
    \delta&= \left[1+ \mu(e^{-D}-1)\right]^N \\ \nonumber
    &= 1+N(e^{-D}-1)+ o(\mu^2[e^{-D}-1]^2) \\ \nonumber
    &= 1+ \mu N(-D) +o(D^2) +o(\mu[e^{-D}-1]^2) \\ \nonumber
    &\approx 1-\mu ND
\end{align}
Similarly, our approximation can be expressed as: 
\begin{equation}
    \delta=e^{-D\mu N} = 1-\mu ND + o([\mu ND]^2)\approx 1-\mu ND
\end{equation}
The advantage of the approximation is that the verifier can draw the conclusion merely based on his received copy numbers $\mu N$, without needing detailed knowledge of the ratio $\mu$ or the total sample number $N$. 

\begin{figure}
\centering
\includegraphics[width=0.8\textwidth]{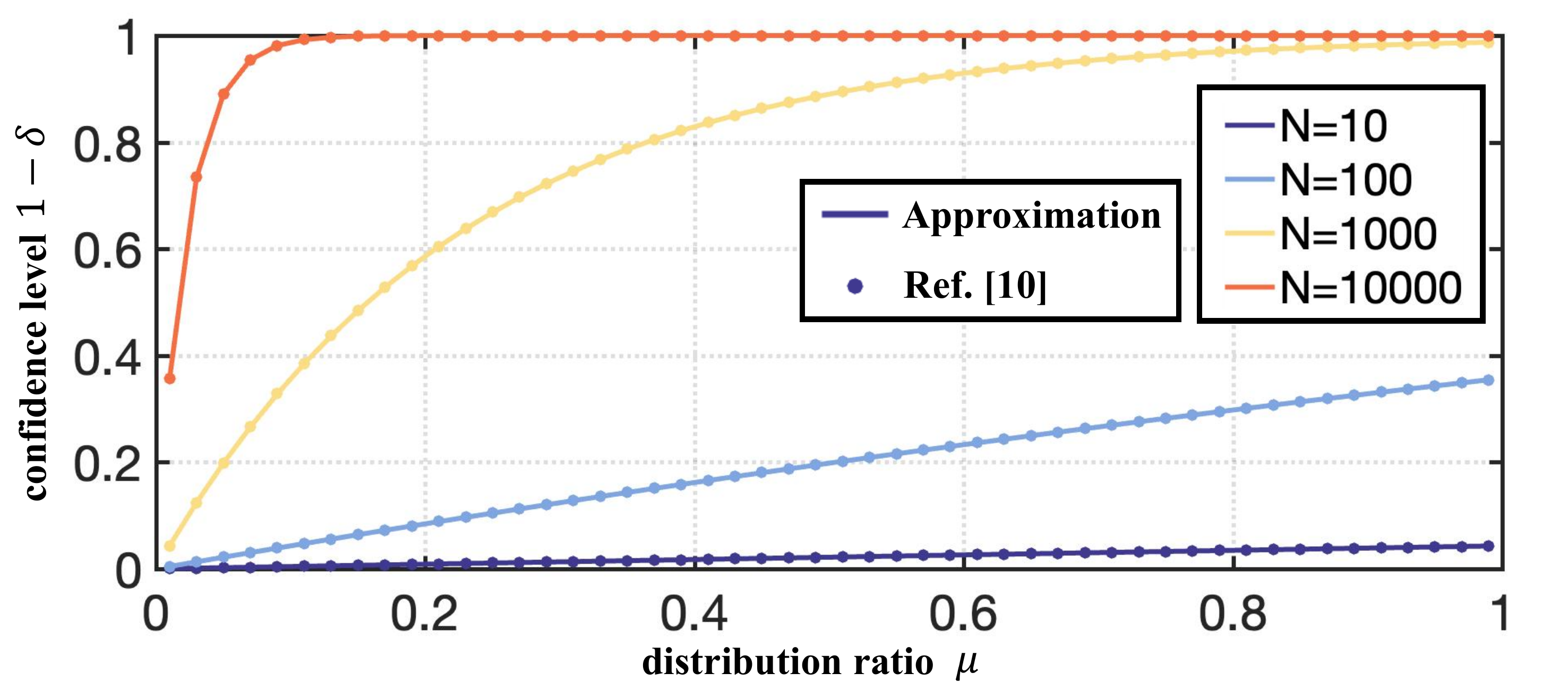}
	\caption{\textbf{comparison of approximate confidence calculation with the original formula.} We adopt the our experimental specification $P_{exp,GHZ}=0.977$ and prefixed $P_\eta=0.96$ aimed to certify. We varied the distribution ratio $\mu$ from 0.01 to 0.99. The lines in our plots represent the results using our approximate formula, while the dots correspond to the results obtained from the original formula of Ref. \cite{gocanin2022}. We analyzed the results for $N=\{10,100,1000,10000\}$. In each case, our approximate formula closely matches the original formula, validating the effectiveness of our approximation for both small and large sample sizes.}
\label{fig:approx}
\end{figure}

We also conducted numerical comparisons of the two definitions under various parameters, as shown in Fig. \ref{fig:approx}. By adopting our experimentally observed success probability $P_{exp,GHZ}=0.977$ and the fidelity to be certified $P_{\eta}=0.96$, we found that the results of the two formulas align well across a distribution ratio  $\mu \in (0,1)$, in both small ($N=10$) and large ($N=10000$) samples regime.

\end{document}